\pdfoutput=1
\documentclass[aps,prresearch,reprint,longbibliography,superscriptaddress]{revtex4-2}

\usepackage{graphicx}
\usepackage{amsmath}
\usepackage{amssymb}
\usepackage{bm}
\usepackage{braket}
\usepackage{xcolor}
\usepackage{comment}
\usepackage[colorlinks=true,linkcolor=blue,citecolor=blue,urlcolor=blue]{hyperref}

\begin{document}
\raggedbottom

\title{Embedding nuclear Hamiltonians and molecules on quantum annealers}

\author{Emanuele Costa}
\email{ecosta@bsc.es}

\author{Alba Cervera Lierta}
\affiliation{Barcelona Supercomputing Center (BSC), Barcelona, Spain}

\date{\today}

\begin{abstract}
Quantum technologies offer new tools to study quantum many-body systems such as atomic nuclei and molecules. Digital quantum computers provide several routes to simulate these systems, but the required gate depth and the sensitivity to noise make such simulations difficult on near-term hardware. Quantum annealers avoid deep circuits and are naturally suited to adiabatic state preparation, which is partially protected against noise by the spectral gap, but their native interactions are too restricted to represent the Hamiltonians that describe these systems. Here we show how to encode pairing-type quasiparticle Hamiltonians, which well describe semi-magic and neutron-rich nuclei in the shell model and molecules within the doubly occupied configuration interaction approximation, into Ising Hamiltonians with local transverse fields. We construct perturbative gadget Ising Hamiltonians that encode the information of the targeted dynamics inside the low-energy spectrum. We benchmark the encoding on oxygen and calcium isotopes and on the LiH molecule, obtaining fidelities with respect to the target state that improve systematically with the strength of the gadget constraint. We then test the gadgets against the practical limitations of real annealers, analysing their minor embedding onto the Pegasus topology of D-Wave devices and the use of a single global transverse field, and we discuss the adiabatic preparation of the nuclear targets. Our results show that simulating molecules and nuclei via gadgets is, in principle, compatible with the native interactions and connectivity of current quantum annealers.
\end{abstract}

\maketitle

\section{Introduction}
\label{sec:intro}

Strongly correlated fermionic systems, such as nucleons in atomic nuclei and electrons in molecules, are among the most natural targets of quantum computation~\cite{RevModPhys.86.153,Ayral:2023ron,Cao2019}. The dimension of their Hilbert space grows exponentially with the number of particles, so that exact diagonalization becomes intractable beyond moderate system sizes, while approximate methods are limited by the fermionic sign problem or by the growth of entanglement~\cite{annurev:/content/journals/10.1146/annurev-nucl-101918-023600,annurev:/content/journals/10.1146/annurev-conmatphys-033117-054307,Orus2019,lange2024architectures}. Quantum simulators and quantum computers avoid this bottleneck by representing many-body states natively, mapping the physical degrees of freedom onto qubits and their dynamics onto controllable Hamiltonians or gate sequences~\cite{Daley2022,10.1098/rspa.1998.0162,Cirac2012QuantumSimulation}. This approach has produced promising results in quantum chemistry~\cite{RevModPhys.92.015003,Bauer2020QuantumAlgorithms,Kandala2017,shajan2024quantumcentricsimulationsextendedmolecules} and nuclear structure~\cite{PhysRevC.106.034325,perez2023nuclear,PhysRevC.105.024324,PhysRevC.110.064320,Bhoy_2024,Li:2023qdd,Sarma:2025ptf,costa2024quantum,17n2-xh6k,Zhang_2021,Ayral:2023ron}.
Most of these demonstrations, however, rely on digital devices, where noise still restricts the accessible circuit depth~\cite{Preskill2018quantumcomputingin,bharti2022noisy}.

A complementary route is offered by analog quantum simulation, proposed also for quantum chemistry~\cite{ArguelloLuengo2019}, and in particular by adiabatic quantum computation and its finite-time, finite-temperature implementation on hardware, quantum annealing (QA)~\cite{Farhi2001,Kadowaki1998,AlbashLidar2018,Hauke2020}. Here, the problem is encoded in a Hamiltonian with user-tunable couplings and the system is slowly driven from the easily prepared ground state of a transverse-field driver to the ground state of the target. For slow enough driving, the spectral gap suppresses transitions out of the instantaneous ground state, which gives adiabatic evolution some robustness against noise~\cite{Childs2001}. Quantum annealers implement this idea with thousands of superconducting qubits~\cite{Johnson2011}. Originally designed for combinatorial optimization~\cite{Lucas2014}, they are increasingly used as analog simulators of quantum magnets, at scales that challenge state-of-the-art classical methods~\cite{Harris2018,King2018,King2023}. It is therefore natural to ask whether annealers can also address fermionic systems.

One of the main challenges for the simulation of quantum many-body systems on annealers is the lack of universality of the controllable Hamiltonians, which are restricted to Ising couplings and local transverse fields~\cite{BravyiStoquastic2008,Biamonte2008,AlbashLidar2018}. To improve the expressibility of the tunable interactions, several works have focused on gadget methods. On the one hand, specific mappings of the logical problem onto the hardware have been designed to overcome the limited connectivity of the Ising couplings~\cite{Lechner2015,Palacios2025}. On the other hand, gadgets have been used to introduce catalyst terms in the adiabatic evolution, which are not natively available on current devices but can improve the performance of the annealing~\cite{Albash2019,Feinstein2024}. Such terms have been generated both with perturbative gadgets~\cite{Banks2025} and with exact, non-perturbative constructions~\cite{Banks2026}.
Furthermore, perturbative gadgets have been applied to the electronic-structure problem~\cite{babbush2014adiabatic}: by adding ancilla qubits and strong penalties, they reduce the $k$-local Hamiltonians produced by the fermion-to-qubit mapping to two-local ones~\cite{KempeKitaevRegev2006,JordanFarhi2008,Bravyi2008}, which still require two-body interactions beyond those of the transverse-field Ising model~\cite{babbush2014adiabatic,Biamonte2008}. Moreover, these reductions require many ancillas per term and penalty scales that grow rapidly with the order of the reduction~\cite{PhysRevA.91.012315}, while the catalyst constructions above apply only to specific coupling topologies.

In this work, we follow a different route and consider quasiparticle Hamiltonians, in which fermions are grouped into pairs that behave as hard-core bosons. This approximation is accurate in relevant regimes both in nuclear structure, where the quasiparticle pairing description reproduces the ground states of semi-magic nuclei~\cite{PhysRevC.109.064305,td9s-z7my,COSTA2026140042,Costa2026BW}, and in quantum chemistry, where it corresponds to the doubly occupied configuration interaction (DOCI) Hamiltonian~\cite{10.1063/1.3613706,Elfving2021,Krompiec:2025fac}. Since each quasiparticle mode is mapped directly onto a spin, the overhead of the fermionic encoding is avoided altogether. We then construct Ising gadget Hamiltonians whose low-energy subspace reproduces the target quasiparticle Hamiltonian through second-order Brillouin-Wigner perturbation theory~\cite{HubacWilson2010,JordanFarhi2008}, in the spirit of the one-hot gadget of Ref.~\cite{Banks2025} but extended to the full network of quasiparticle hoppings, and we compare two encodings of the logical space. In contrast to general-purpose constructions, the gadget requires only the native Ising couplings and transverse fields of current annealers and no ancilla qubits per interaction term, while adjustable virtual-state denominators and explicit counterterms make the effective Hamiltonian directly inspectable and allow us to quantify the trade-off between accuracy and hardware resources. Gadget constructions are usually analysed at the level of the effective Hamiltonian, assuming arbitrary connectivity and independently tunable local fields. Here we go one step further and test our gadgets against the constraints of a real annealer. We embed them onto the Pegasus topology of commercial D-Wave devices~\cite{Boothby2020}, where the chains renormalize the effective transverse fields and bound the usable penalty strength from above, and we show that a single global transverse field can replace the local ones for stoquastic targets, at the cost of a finite-resource penalty. Finally, we study adiabatic state preparation of the nuclear targets.

\begin{figure*}[ht!]
    \centering
    \includegraphics[width=\textwidth]{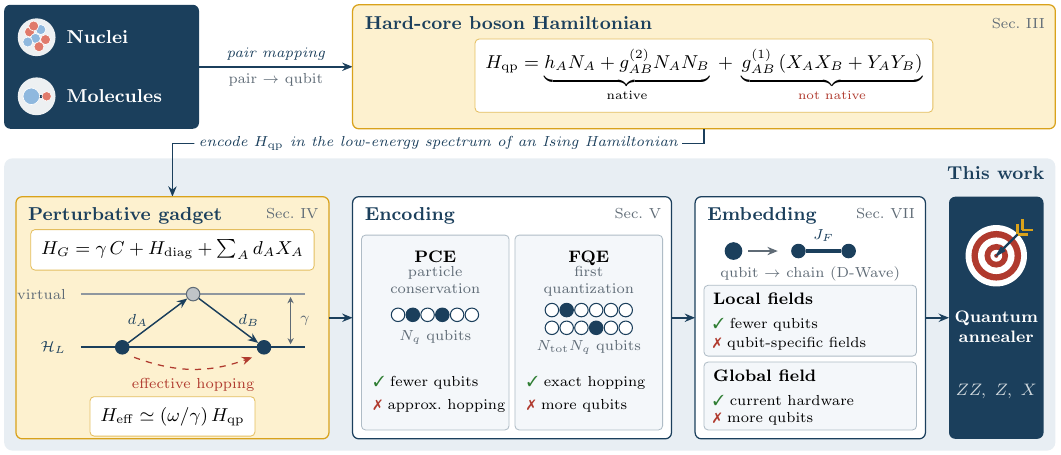}
    \caption{Overview of this work. Our goal is to encode fermionic systems, in particular nuclei and molecules, into a Hamiltonian that can be simulated with a quantum annealer. To do so, we use perturbative gadgets that encode the hopping terms from the hard-core boson Hamiltonians of these systems into the native interactions of an annealer. We use two encodings: the particle conservation encoding (that requires less qubits at the cost of approximating the hopping term) and the first quantization encoding (that requires more qubits but matches the hopping term exactly at second order for admissible couplings). Finally, we explore two embedding methods that translate the final Hamiltonian into a realistic quantum annealing one, where the qubit connectivity is fixed and the transverse fields are mostly global.}
    \label{fig:overview}
\end{figure*}

\section{Overview}
\label{sec:overview}

Figure~\ref{fig:overview} summarizes the methodology used in this work. We start by describing the physical systems of interest, strongly correlated fermionic systems: nuclei and molecules. When pairing correlations dominate, the fermionic Hamiltonian is projected onto configurations of time-reversed pairs, which behave as hard-core bosons and map one-to-one onto qubits. Both semi-magic nuclei (Sec.~\ref{sec:nsm_qp_hamiltonian}) and molecules in the DOCI approximation (Sec.~\ref{sec:molecular_hamiltonian}) take the same form, 
\begin{multline}
  H_\text{qp}=\sum_A h_AN_A+\sum_{A<B}g^{(2)}_{AB}N_AN_B \\ +\sum_{A<B}g^{(1)}_{AB}\left(X_AX_B+Y_AY_B\right),
  \label{eq:H_qp}
\end{multline}
where $N\equiv(1 - Z)/2$ is the number operator and $X$, $Y$, $Z$ are the Pauli matrices acting on the corresponding pairs. The coupling terms correspond to the on-site energies $h_A$, density--density couplings $g^{(2)}_{AB}$ and pair hopping $g^{(1)}_{AB}$, all in units of an energy scale $\omega$ ($1$~MeV for nuclei, $1$~Ha for the molecule LiH). While the diagonal terms $N_{A}N_{B}$ are native to an annealer, the hopping term is not. The goal of this work is to transform that term and find the proper qubit embedding so that these systems can be mapped onto the native interactions and connectivity of current quantum annealers.

\emph{Gadget} (Sec.~\ref{sec:gadget}). A penalty $\gamma C$, built from $N_A$ and $N_AN_B$ terms, vanishes on the physical configurations (so-called the logical space $\mathcal{H}_L$), and costs at least $\gamma$ elsewhere. A transverse field $\sum_Ad_AX_A$ with $|d_A|\ll\gamma$ leaves $\mathcal{H}_L$ with one flip and returns with a second one, moving a pair from $A$ to $B$. At second order this gives
\begin{equation}
  H_\text{eff}=[ H_\text{diag}-\frac{1}{2\gamma}\sum_{A<B}d_Ad_B\Delta_{AB}\left(X_AX_B+Y_AY_B\right)],
  \label{eq:Heff_full}
\end{equation}
where $H_\text{diag}$ collects the diagonal target terms, programmed directly, and $\Delta_{AB}>0$ is set by the energies of the two virtual states of the hop. Two consequences recur throughout the paper. The target is reproduced up to a factor $\omega/\gamma$, which compresses the logical spectrum to the scale $\omega^2/\gamma$, and the generated hopping has the sign of $-d_Ad_B$, so the gadget reproduces stoquastic Hamiltonians.

\emph{Encoding} (Sec.~\ref{sec:encodings}). The penalty $C$ defines the logical space and, through the virtual energies, $\Delta_{AB}$. The particle conservation encoding (PCE, Sec.~\ref{sec:pce}) fixes the number of pairs on $N_q$ qubits. It imposes $\Delta_{AB}=2$, hence a rank-one approximation of the hopping. The first quantization encoding (FQE, Sec.~\ref{sec:fqe}) assigns a one-hot register to each pair, using $N_\text{tot}N_q$ qubits. Then, intra-register couplings tune $\Delta_{AB}$ pair by pair and match the hopping. 

\emph{Benchmark} (Section~\ref{sec:results}). We test both encodings on O and Ca isotopes (Sec.~\ref{sec:results_nsm}) and on LiH (Sec.~\ref{sec:results_lih}). The frustrated DOCI hopping requires there either a sign reversal or a stoquastic approximation (App.~\ref{app:stoq}). Section~\ref{sec:qa} prepares the nuclear ground states using adiabatic state preparation. While PCE requires less qubits than FQE and represents an approximation of the hopping terms, it obtains good fidelities with respect to the exact ground state, for low values of $\gamma$. On the other side, the FQE reproduces the hopping exactly at second order for admissible couplings (Sec.~\ref{sec:fqe}), at the cost of using more qubits and larger penalties $\gamma$.

\emph{Embedding} (Sec.~\ref{sec:embedding}). On the Pegasus topology (D-Wave quantum annealers) each gadget qubit becomes a ferromagnetic chain of strength $J_F$, which acts as a single qubit with a renormalized drive (Sec.~\ref{sec:minor_embedding}, App.~\ref{app:chain}). While this embedding is minimal, it requires custom transverse field for each physical qubit. On the other hand, we propose another embedding that uses a single transverse field, which matches the global-drive control of current hardware (Sec.~\ref{sec:global_field}) and find equivalent results at the cost of using more qubits.

\section{Quasiparticle Hamiltonians}
\label{sec:theory}

Strongly interacting fermionic systems, from atomic nuclei to molecules, are
commonly described within a finite single-particle basis by Hamiltonians
containing one- and two-body terms~\cite{RevModPhys.77.427,doi:10.1021/acs.jctc.3c00123,RevModPhys.92.015003}.
When pairing correlations dominate~\cite{RevModPhys.75.607}, the low-energy physics is captured to a good
approximation by configurations in which fermions occupy
time-reversed pairs of orbitals, i.e., $(m,-m)$ pairs in nuclei and doubly occupied spatial orbitals in molecules~\cite{talmi_book,10.1063/1.3613706}.
Each pair behaves as a hardcore boson, so that the problem maps onto a spin-$1/2$
model without fermionic sign strings~\cite{Elfving2021,PhysRevC.105.024324,Krompiec:2025fac}.
In this work we exploit this structure for nuclei in the shell model and,
with the same formalism, for seniority-zero molecular
Hamiltonians.

\subsection{The quasiparticle pairing Hamiltonian}
\label{sec:nsm_qp_hamiltonian}

A nucleus with valence nucleons in an active space is described by the
nuclear shell model (NSM) Hamiltonian as
\begin{equation}
    H_\text{NSM}=\sum_a e_a\, c^{\dagger}_a c_a
    + \frac{1}{4}\sum_{abcd} v_{abcd}\, c^{\dagger}_a c^{\dagger}_b c_d c_c ,
    \label{eq:NSM}
\end{equation}
where $c^{\dagger}_a$ ($c_a$) creates (annihilates) a nucleon in mode $a$,
$e_a$ are the single-particle energies and $v_{abcd}$ the antisymmetrized
two-body matrix elements. The latter can be derived from chiral
interactions~\cite{Hjorth-Jensen:1995zrg,Coraggio:2008in,annurev:/content/journals/10.1146/annurev-nucl-101917-021120,PhysRevC.91.064301,PhysRevLett.113.142502}
or adjusted phenomenologically to reproduce experimental
data~\cite{RevModPhys.77.427,RevModPhys.92.015002}. 

The restriction to a
valence space is the nuclear analogue of the active-space methods of quantum
chemistry~\cite{doi:10.1021/acs.jctc.3c00123}.
Each mode labels a single-particle state
$\ket{a}=\ket{n_a,l_a,j_a,m_a,t_{z,a}}$, with $n_a$ the radial quantum number,
$l_a$ the orbital angular momentum, $j_a$ the total angular momentum, $m_a$ its
projection, and $t_{z,a}=\pm1/2$ the isospin projection distinguishing protons
from neutrons. Orbitals, defined as sets of modes differing only in $m_a$, are
denoted in spectroscopic notation $n l_j$, and for each nucleon species the
number of modes is $D=\sum_{j\in J}(2j+1)$, with $J$ the set of valence
orbitals. The many-body basis is spanned by Slater determinants
$\ket{\mathbf{s}}=\prod_{a:\,s_a=1}c^{\dagger}_a\ket{0}$, labelled by bitstrings
$\mathbf{s}\in\{0,1\}^{2D}$, so that for $Z_p$ valence protons and $N_n$
valence neutrons $\dim\mathcal{H}=\binom{D}{Z_p}\binom{D}{N_n}$.
We consider two valence spaces: the $sd$ shell
($0d_{5/2},1s_{1/2},0d_{3/2}$; $D=12$) with the USDB
interaction~\cite{PhysRevC.74.034315}, and the $pf$ shell
($0f_{7/2},1p_{3/2},0f_{5/2},1p_{1/2}$; $D=20$) with
GXPF1A interaction~\cite{PhysRevC.69.034335,Honma2005}.

The complexity of the fermionic problem is reduced by working with pairs of
nucleons in single-particle states with opposite angular-momentum
projection~\cite{COSTA2026140042,PhysRevC.109.064305,td9s-z7my},
created by the quasiparticle operators
\begin{equation}
    Q^{\dagger}_A = c^{\dagger}_{a}\, c^{\dagger}_{\tilde{a}}, \qquad
    N_A = Q^{\dagger}_A Q_A ,
\end{equation}
where $\tilde{a}$ is the partner of $a$ with opposite projection,
$\ket{\tilde a}=\ket{n_a,l_a,j_a,-m_a,t_{z,a}}$. The index
$A\equiv\{n_a,l_a,j_a,t_{z,a},|m_a|\}$ runs over $m_a>0$ only, so that a $j$
orbital hosts $(2j+1)/2$ pair modes and each species $D/2$ of them.
These operators obey a hardcore-boson
algebra~\cite{doi:10.1142/S0217979206034947},
\begin{align}
    [Q_A,Q^{\dagger}_B] &= \delta_{AB}\,(1-n_{a}-n_{\tilde{a}}), \nonumber \\
    [Q_A,Q_B] &= [Q^{\dagger}_A,Q^{\dagger}_B]=0, \qquad \big(Q^{\dagger}_A\big)^2=0 ,
    \label{eq:hb_algebra}
\end{align}
i.e.\ pairs on different modes commute, while double occupancy of the same mode
is excluded. On the $(m,-m)$ pair subspace, where $n_a=n_{\tilde a}=N_A$, the
first relation becomes $[Q_A,Q^\dagger_B]=\delta_{AB}(1-2N_A)$, which is
reproduced exactly by the Pauli algebra
\begin{equation}
\begin{aligned}
    Q^{\dagger}_A &\mapsto \sigma^{-}_A=\tfrac{1}{2}(X_A-iY_A), \\
    Q_A &\mapsto \sigma^{+}_A=\tfrac{1}{2}(X_A+iY_A), \\
    N_A &\mapsto \tfrac{1}{2}(1-Z_A),
\end{aligned}
\label{eq:pauli_map}
\end{equation}
with $\ket{1}_A$ ($Z_A=-1$) denoting an occupied pair. Since pairs on different
modes commute, the encoding requires no Jordan--Wigner~\cite{10.21468/SciPostPhysLectNotes.82}
or Bravyi--Kitaev~\cite{bk,Seeley2012} strings, which are used in most
shell-model quantum simulations to date~\cite{PhysRevC.105.064317,perez2023nuclear,Bhoy_2024,17n2-xh6k}.

Projecting Eq.~\eqref{eq:NSM} onto the $(m,-m)$ pair subspace of like
nucleons, $H_\text{NSM-Q}=\mathcal{P}_0 H_\text{CI}\mathcal{P}_0$~\cite{COSTA2026140042}, and using
Eq.~\eqref{eq:pauli_map}, we obtain
\begin{equation}
H_{\text{NSM-Q}} = H_\text{qp}
\label{eq:H_NSMQ}
\end{equation}
from Eq.~\eqref{eq:H_qp} with
\begin{equation}
\begin{split}
h_A           &= 2e_a + v_{a\tilde a a\tilde a}, \\
g^{(1)}_{AB}  &= \tfrac{1}{2}\, v_{a\tilde a b\tilde b}, \\
g^{(2)}_{AB}  &= v_{abab} + v_{\tilde a b\tilde a b} + v_{a\tilde b a\tilde b}
                + v_{\tilde a\tilde b\tilde a\tilde b}.
\end{split}
\label{eq:coeffs}
\end{equation}
Here $h_A$ is the energy of an isolated pair, given by twice the
single-particle energy ($e_a=e_{\tilde a}$ by rotational invariance) plus the
diagonal pairing matrix element; $g^{(1)}_{AB}$ is the pair-hopping amplitude
generated by transitions of $M=0$ pairs; and $g^{(2)}_{AB}$ is the
density--density interaction between two occupied pairs, which collects the
direct, monopole-like part of $v$. All couplings of the target Hamiltonians are expressed in units of an energy scale $\omega=1$~MeV. All terms of $H_\text{CI}$ that break a pair,
i.e.\ that connect $\mathcal{P}_0\mathcal{H}$ to states with at least one broken pair, are neglected; their leading effect can be reinstated
perturbatively, e.g.\ within Brillouin--Wigner theory~\cite{HubacWilson2010,Costa2026BW}.

The mapping halves the number of qubits per species, $N_q=D/2$, i.e.\
$N_q=6$ and $10$ in the $sd$ and $pf$ shells, and reduces the Hilbert space to
$\binom{D/2}{N_n/2}$, the number of ways to distribute $N_n/2$ neutron pairs.
The approximation is accurate for semi-magic nuclei, where only one nucleon species is active and configurations of $(m,-m)$ pairs dominate the ground state~\cite{talmi_book,PhysRevC.109.064305}, which $H_\text{NSM-Q}$ reproduces with $|\langle\Psi_\text{NSM-Q}|\Psi_\text{CI}\rangle|^2\geq0.95$~\cite{COSTA2026140042}. We therefore focus on the
oxygen ($sd$ shell) and calcium ($pf$ shell) isotopes, both with
$Z_p=0$.

The couplings for both valence spaces are shown in Fig.~\ref{fig:couplings} of Appendix~\ref{app:couplings}. Both
$g^{(1)}_{AB}$ and $g^{(2)}_{AB}$ connect every pair of modes, and this
all-to-all structure is the main obstacle for hardware with local
connectivity. The signs of the couplings, instead, are less of an obstacle
than they appear. The hopping amplitudes alternate in sign
[Figs.~\ref{fig:couplings}(a) and \ref{fig:couplings}(c)], but this pattern
depends on the phase convention of the pairs: exchanging the order of the two
fermionic operators in a pair ($a\leftrightarrow\tilde a$, with
$c^\dagger_{\tilde a}c^\dagger_a=-c^\dagger_a c^\dagger_{\tilde a}$) maps
$Q_A\to s_A Q_A$, $s_A=\pm1$, and hence
$g^{(1)}_{AB}\to s_As_B\,g^{(1)}_{AB}$, leaving the spectrum and all
occupations unchanged. In both valence spaces a choice of $\{s_A\}$ exists that
makes every hopping amplitude non-positive, since no triangle of modes is
frustrated ($g^{(1)}_{AB}g^{(1)}_{BC}g^{(1)}_{CA}<0$)~\cite{Harary1953,Hen2021}.
As $h_A$ and $g^{(2)}_{AB}$ are diagonal, $H_\text{NSM-Q}$ is therefore
stoquastic~\cite{BravyiStoquastic2008}. As shown in Sec.~\ref{sec:gadget},
this is what makes the gadget easy to apply for this class of Hamiltonians.

The density--density couplings behave differently in the two valence spaces.
For USDB the large couplings are all attractive,
$g^{(2)}_{AB}\in[-2.30,-0.17]$~MeV, while the few repulsive ones are smaller
than $0.06$~MeV. For GXPF1A, instead, attractive and repulsive couplings of
comparable magnitude coexist, $g^{(2)}_{AB}\in[-1.84,+1.79]$~MeV. 

Finally, orbitals with $j=1/2$ host a single pair, whose density is
rotationally invariant and couples to the other pairs only through the
monopole. As a result, the $1s_{1/2}$ row of $g^{(2)}$ is constant within each
orbital ($+0.029$~MeV towards $0d_{5/2}$, $+0.015$~MeV towards $0d_{3/2}$), and
so is the $1p_{1/2}$ row in the $pf$ shell ($+0.55$, $-0.97$ and $+0.30$~MeV
towards $0f_{7/2}$, $1p_{3/2}$ and $0f_{5/2}$); the same holds for
$|g^{(1)}|$. In USDB the $1s_{1/2}$ pair is thus almost decoupled in the
density channel and interacts with the other modes mainly through pair
hopping.

\subsection{The DOCI Hamiltonian in LiH}
\label{sec:molecular_hamiltonian}

In quantum chemistry, the electronic structure of a molecule is described in a
finite basis set, such as the minimal STO-3G
basis~\cite{Hehre1969}.
A restricted Hartree--Fock (RHF) calculation in this basis provides orthonormal molecular
orbitals $\phi_\alpha$, in which the active-space
Hamiltonian reads~\cite{Helgaker2000}
\begin{multline}
    H_{\text{FCI}} = E_{\text{nucl}}
    + \sum_{\alpha\beta}\sum_{\sigma}\tilde h^{(1)}_{\alpha\beta}\,
      c^\dagger_{\alpha\sigma} c_{\beta\sigma} \\
    + \frac12\sum_{\alpha\beta\gamma\delta}\sum_{\sigma\tau}
      \tilde h^{(2)}_{\alpha\beta\gamma\delta}\,
      c^\dagger_{\alpha\sigma} c^\dagger_{\beta\tau} c_{\gamma\tau} c_{\delta\sigma},
\label{eq:H_FCI}
\end{multline}
where $\sigma,\tau\in\{\uparrow,\downarrow\}$,
\begin{equation}
    \tilde h^{(2)}_{\alpha\beta\gamma\delta}
    = \int d\mathbf r_1\, d\mathbf r_2\,
      \frac{\phi^*_\alpha(\mathbf r_1)\phi^*_\beta(\mathbf r_2)\,
            \phi_\delta(\mathbf r_1)\phi_\gamma(\mathbf r_2)}
           {|\mathbf r_1-\mathbf r_2|}
\end{equation}
are the electron--electron integrals, \begin{equation} \tilde h^{(1)}_{\alpha\beta} =\int d\mathbf r\,\phi^*_\alpha(\mathbf r)\Big[-\frac{\nabla^2}{2}-\sum_I\frac{Z_I}{|\mathbf r-\mathbf R_I|}\Big]\phi_\beta(\mathbf r) \end{equation} are the one-electron integrals, which contain the kinetic energy and the attraction to the nuclei of charge $Z_I$ at positions $\mathbf R_I$, and $E_{\text{nucl}}$ is the nuclear repulsion energy.
The many-body basis is spanned by the Slater determinants
$\ket{\mathbf s}=\prod_{(\alpha\sigma):\,s_{\alpha\sigma}=1}c^\dagger_{\alpha\sigma}\ket{0}$.

In analogy with the nuclei Hamiltonian from the previous section, $H_\text{NSM-Q}$, we introduce the pair operators
$Q^\dagger_A=c^\dagger_{\alpha\uparrow}c^\dagger_{\alpha\downarrow}$, which create
two electrons with opposite spin in the spatial orbital $\alpha$ (labelled by the
pair mode $A$) and obey the same hardcore-boson algebra from Eq.~\eqref{eq:hb_algebra}. Projecting
Eq.~\eqref{eq:H_FCI} onto the seniority-zero subspace, i.e.\ discarding
single-electron hopping and spin-flip terms, yields the doubly occupied
configuration interaction (DOCI) Hamiltonian~\cite{10.1063/1.3613706,Elfving2021}.
In the qubit representation of Eq.~\eqref{eq:pauli_map} it reads
\begin{equation}
H_{\text{DOCI}} = H_{\text{qp}} + E_{\text{nucl}} 
\label{eq:H_DOCI}
\end{equation}
i.e.\ it has the same form as Eq.~\eqref{eq:H_qp}, up to the constant $E_{\text{nucl}}$, with
\begin{align}
    h_A &= 2\tilde h^{(1)}_{\alpha\alpha} + \tilde h^{(2)}_{\alpha\alpha\alpha\alpha}, \nonumber\\
    g^{(1)}_{AB} &= \tfrac12\,\tilde h^{(2)}_{\alpha\alpha\beta\beta}, \label{eq:doci_coeffs}\\
    g^{(2)}_{AB} &= 2\left(2\,\tilde h^{(2)}_{\alpha\beta\beta\alpha} - \tilde h^{(2)}_{\alpha\beta\alpha\beta}\right). \nonumber
\end{align}
For real orbitals $\tilde h^{(2)}_{\alpha\alpha\beta\beta}$ is an exchange
integral, so $g^{(1)}_{AB}>0$ for every pair $A,B$.
The interaction term $h_A$ is
the energy of a doubly occupied orbital: twice its one-body energy plus the
Coulomb repulsion $\tilde h^{(2)}_{\alpha\alpha\alpha\alpha}$ between its two
electrons. The term $g^{(1)}_{AB}$ transfers an electron pair from orbital $\beta$ to
orbital $\alpha$ through the exchange integral. Finally, the term $g^{(2)}_{AB}$ is the
interaction between two doubly occupied orbitals: the direct Coulomb
repulsion among the four electrons, $4\tilde h^{(2)}_{\alpha\beta\beta\alpha}$,
reduced by the exchange between the two pairs of parallel spins,
$2\tilde h^{(2)}_{\alpha\beta\alpha\beta}$. The energy units for LiH are $\omega=1$~Ha.

Following Ref.~\cite{Elfving2021}, we take LiH as molecular test case, in the
STO-3G basis at the experimental equilibrium bond length
$R=1.595$~\AA~\cite{PearsonGordy1969}. The RHF orbitals and the integrals entering
Eq.~\eqref{eq:doci_coeffs} are computed with PySCF~\cite{PySCF2018,PySCF2020}. The six RHF orbitals,
occupied by two electron pairs, give $N_q=6$ qubits and a seniority-zero space
of dimension $\binom{6}{2}=15$. The resulting couplings are shown in Fig.~\ref{fig:couplings_LiH} of Appendix~\ref{app:couplings}. As in the nuclear case, both $g^{(1)}_{AB}$ and
$g^{(2)}_{AB}$ connect all pairs of modes, but their structure is markedly
different. The pair hopping is small, $g^{(1)}_{AB}\in[0.005,0.063]$~Ha, and
strictly positive, so every triangle of modes is frustrated,
$g^{(1)}_{AB}g^{(1)}_{BC}g^{(1)}_{CA}>0$: no choice of $\{s_A\}$ makes all the
hopping amplitudes non-positive, and $H_\text{DOCI}$ is not stoquastic in the
pair basis. The density--density couplings, instead, are large and purely
repulsive, $g^{(2)}_{AB}\in[0.87,1.58]$~Ha, two orders of magnitude above the
hopping, so no energy shift is needed on platforms with repulsive interactions.

\section{Ising Gadget Hamiltonian}
\label{sec:gadget}

\subsection{Annealer Hamiltonian}
\label{sec:annealer_hamiltonian}

Quantum annealers are a technology based on the adiabatic evolution of a time-dependent Hamiltonian $H(t)$ with a fixed structure. The native Hamiltonians that can be implemented on these machines belong to the Ising class
\begin{equation}
    H_G = H_Z + H_X,
    \label{eq:HG}
\end{equation}
with longitudinal and transverse parts
\begin{equation}
    H_Z = \sum_{i<j}^{N_G} J_{ij} Z_i Z_j + \sum_{i}^{N_G} h_i Z_i,
    \qquad
    H_X = \sum_{i}^{N_G} d_i X_i,
    \label{eq:HZHX}
\end{equation}
where $N_G$ is the total number of qubits. 
For the sake of generality, we let the couplings $J_{ij}$
vary in both connectivity and value, as well as the transverse fields $d_{i}$ and $h_{i}$.
The $H_Z$ terms are diagonal in the computational basis, while $H_{X}$ flips one qubit at a time. Since the logical states will be quasiparticle occupations, it is convenient to write the diagonal part in terms of the number operators $N_i = \tfrac{1}{2}\left(1 - Z_i\right)$.

\subsection{Gadget construction}
\label{sec:gadget_construction}

The annealer Hamiltonian from Eq.~\eqref{eq:HG} cannot represent directly the quasiparticle systems described in the previous section (Eq.~\eqref{eq:H_qp}), since it contains no hopping term $X_AX_B+Y_AY_B$. On the other hand, if its Hilbert space $\mathcal{H}$ is much larger than the space of the system we want to describe, confined in a logical space $\mathcal{H}_L$, the dynamics of interest can be encoded in a part of the spectrum of $H_G$ as an effective dynamics.

Using the Brillouin--Wigner formalism~\cite{Lowdin1962,HubacWilson2010}, we build into $H_Z$ a constraint term $C$, made only of $N_i$ and $N_iN_j$ operators, that energetically separates the logical subspace $\mathcal{H}_L$ from the rest of the Hilbert space. We denote by $\gamma$ the energy scale of this separation and by $\omega$ the energy unit of the target couplings (fixed in Sec.~\ref{sec:theory} for each system). We then choose the couplings $J_{ij}$ and $h_i$ of Eq.~\eqref{eq:HZHX} so that the longitudinal part splits as
\begin{equation}
    H_Z = \gamma C + H_{\text{diag}}, \qquad H_G = \gamma C + H_{\text{diag}} + H_X,
    \label{eq:HG_C}
\end{equation}
where $C$ is a dimensionless operator and $H_{\text{diag}}$, of order $\omega^2/\gamma$, collects the diagonal target terms. Since $\gamma\gg\omega$, we expand around $\gamma C$ alone and treat both $H_{\text{diag}}$ and the transverse part $H_X$, of order $d=\max_i|d_i|=O(\omega)$ with $d_i$ the transverse fields of Eq.~\eqref{eq:HZHX}, as perturbations.

We then define $\mathbb{L}$ as the projector onto the logical space, so we can construct an effective dynamics inside that space governed by $H_G$ by using the effective Hamiltonian
\begin{equation}
    H_{\text{eff}}(E)=H_{\text{LL}}+ H_{\text{LR}}(E-H_{\text{RR}})^{-1}H_{\text{RL}},
    \label{eq:BW}
\end{equation}
where $H_{\text{LL}}=\mathbb{L}H_G\mathbb{L}$, $H_{\text{LR}}=\mathbb{L}H_G\mathbb{R}$, $H_{\text{RL}}=\mathbb{R}H_G\mathbb{L}$ and $H_{\text{RR}}=\mathbb{R}H_G\mathbb{R}$, with $\mathbb{R}=\mathbb{I}-\mathbb{L}$ the projector onto the complementary space. In the logical subspace we have that
\begin{equation}
    H_{\text{LL}}=\mathbb{L} H_{\text{diag}} \mathbb{L},
\end{equation}
because $H_X$ does not couple logical states and $C$ vanishes on the logical subspace. 

Equation~\eqref{eq:BW} is exact, but it has to be solved self-consistently because the eigenvalue $E$ enters the resolvent. We take the constraint to be non-negative and to vanish exactly on the logical subspace, so that the unperturbed logical energy is $E_0=0$ and the virtual states lie at energies of order $\gamma$. 

To expand Eq.~\eqref{eq:BW} in powers of $d/\gamma$, we split the virtual block as $H_{\text{RR}}=D_{\text{RR}}+V_{\text{RR}}$, where $D_{\text{RR}}=\mathbb{R}(\gamma C+H_{\text{diag}})\mathbb{R}$ is diagonal in the computational basis and $V_{\text{RR}}=\mathbb{R}H_X\mathbb{R}$ is the transverse field coupling different virtual states. If we expand the resolvent~\cite{KempeKitaevRegev2006} in terms of $V_{\text{RR}}$ and $G_0\equiv(E_0-D_{\text{RR}})^{-1}=O(\gamma^{-1})$, the unperturbed resolvent, we get
\begin{equation}
  \begin{split}
    (E_0-H_{\text{RR}})^{-1} &= G_0 + G_0 V_{\text{RR}} G_0 \\
    &\quad + G_0 V_{\text{RR}} G_0 V_{\text{RR}} G_0 + \dots,
  \end{split}
  \label{eq:resolvent}
\end{equation}
in which each additional factor $V_{\text{RR}}G_0$ carries one power of $d/\gamma$. The replacement $E\to E_0$ is consistent with this expansion: since $E-E_0=O(d^2/\gamma)$, expanding $(E-H_{\text{RR}})^{-1}$ around $E_0$ we obtain
\begin{equation}
  H_{\text{eff}} = H_{\text{LL}} + H_{\text{LR}}\,G_0\,H_{\text{RL}} + O\!\left(d^4/\gamma^3\right).
  \label{eq:Heff2}
\end{equation}
Each transition of order $d^3/\gamma^2$ involves three single-spin flips, so it cannot bring a logical state back into $\mathcal{H}_L$. The first surviving corrections, of order $d^4/\gamma^3$, come from the term with two factors $V_{\text{RR}}$ and from the energy dependence of $G_0$. Eq \eqref{eq:Heff2} represents the second-order Brillouin-Wigner effective Hamiltonian~\cite{Lowdin1962,HubacWilson2010}. The term $\mathbb{R}H_{\text{diag}}\mathbb{R}=O(\omega^2/\gamma)$ in $D_{\text{RR}}$ shifts the virtual energies only by a relative $O(\omega^2/\gamma^2)$, and therefore contributes at $O(\omega^4/\gamma^3)$, within the error of Eq.~\eqref{eq:Heff2}; in the following we evaluate $G_0$ with $D_{\text{RR}}\simeq\gamma\,\mathbb{R}C\mathbb{R}$.

The structure of the second-order term follows from the locality of the drive. Because $H_X$ is a sum of single-qubit flips, $H_{\text{RL}}$ maps each logical state onto virtual states at Hamming distance one, and $H_{\text{LR}}G_0H_{\text{RL}}$ therefore connects logical states at Hamming distance zero or two. Processes at distance zero flip the same qubit twice and renormalize the diagonal energies. This self-energy depends on the encoding and is discussed in the next section. Processes at distance two generate the off-diagonal couplings. If the constraint fixes the number of excitations, two logical states at Hamming distance two differ by moving one excitation from qubit $i$ to qubit $j$. The off-diagonal part of Eq.~\eqref{eq:Heff2} then reads
\begin{equation}
  \begin{split}
    \left.H_{\text{LR}}\,G_0\,H_{\text{RL}}\right|_{\text{off}} = &-\frac{1}{2\gamma}\sum_{i<j} d_i d_j\,\Delta_{ij}\\
    &\times\mathbb{L}\left(X_i X_j + Y_i Y_j\right)\mathbb{L},
  \end{split}
  \label{eq:Heff_offdiag}
\end{equation}
with the dimensionless weight
\begin{equation}
  \Delta_{ij} \equiv \gamma\sum_{v\in\mathcal{V}_{ij}}\frac{1}{E_v-E_0}.
  \label{eq:DeltaAB}
\end{equation}
Here $\mathcal{V}_{ij}$ contains the virtual states through which the excitation moves, reached by emptying $i$ or by filling $j$ first, and $E_v$ is the corresponding eigenvalue of $D_{\text{RR}}$. Since $E_v-E_0>0$, every $\Delta_{ij}$ is positive and the generated hopping has the sign of $-d_id_j$. 

We consider two constraints, tailored to two encodings of the quasiparticle Hamiltonians: the particle-conservation encoding (PCE, Sec.~\ref{sec:pce}) and the first-quantization encoding (FQE, Sec.~\ref{sec:fqe}).

\section{Gadget constraints from quasiparticle encodings} \label{sec:encodings}

\subsection{Particle Conservation Encoding (PCE)}
\label{sec:pce}

The quasiparticle Hamiltonians of Sec.~\ref{sec:theory} conserve the total number of quasiparticles $N_{\text{tot}}$, whether these are nucleon pairs or electron pairs. A natural way to single out the logical space $\mathcal{H}_L$ is therefore to penalise any deviation from $N_{\text{tot}}$,
\begin{equation}
  \begin{split}
    C = &\Big(\sum_i N_i - N_{\text{tot}}\Big)^2 \\
      &= 2\sum_{i<j} N_i N_j + \left(1-2N_{\text{tot}}\right)\sum_i N_i + N_{\text{tot}}^2,
  \end{split}
  \label{eq:C_PC}
\end{equation}
where we used $N_i^2=N_i$. This encoding uses one qubit per orbital, with $N_G=N_q$, and the constraint is native to the device, at the price of a uniform all-to-all coupling $2\gamma N_iN_j$ between every pair of qubits. Each qubit encodes one quasiparticle mode, $i\equiv A$. The diagonal terms $h_AN_A$ and $g^{(2)}_{AB}N_AN_B$ of Eq.~\eqref{eq:H_qp}, which must enter at the same order $\omega^2/\gamma$ as the hopping, are then carried by
\begin{equation}
  \begin{split}
    H_{\text{diag}} = \frac{\omega}{\gamma}\Big[ &\sum_A \left(h_A+\theta_A\right) N_A \\
    &+ \sum_{A<B}\left(g^{(2)}_{AB}+\theta_{AB}\right) N_A N_B \Big],
  \end{split}
  \label{eq:HZ}
\end{equation}
where $h_A$ and $g^{(2)}_{AB}$ are expressed in units of the energy scale $\omega$, and $\theta_A$, $\theta_{AB}$ are counterterms chosen to cancel the self-energy of the distance-zero processes. Eq.~\eqref{eq:HZ} enters directly in the longitudinal term of the Ising Hamiltonian of Eq.~\eqref{eq:HZHX}.

Every logical state is an eigenstate of $\sum_A N_A$ with eigenvalue $N_{\text{tot}}$, so $C=0$ on $\mathcal{H}_L$ and $E_0=0$. A single flip changes the particle number by $\pm1$, so every virtual state reached by $H_{\text{RL}}$ lies in a sector $N_{\text{tot}}\pm1$, where $C=1$ and $E_v=\gamma$ independently of the occupation of the other orbitals. A hop from an occupied orbital $A$ to an empty orbital $B$ proceeds through two virtual states: one in the $N_{\text{tot}}-1$ sector, reached by removing the quasiparticle from $A$ first, and one in the $N_{\text{tot}}+1$ sector, reached by adding a quasiparticle to $B$ first. Equation~\eqref{eq:DeltaAB} then gives
\begin{equation}
  \Delta_{AB} = 2 \qquad \text{for all } A\neq B .
\end{equation}
No hop can end on an occupied orbital, since a qubit cannot be doubly occupied, so the hard-core condition of Eq.~\eqref{eq:H_qp} is built into the encoding. 

With $\Delta_{AB}=2$, the effective Hamiltonian coincides, up to a constant, with $(\omega/\gamma)\,H_{\text{qp}}$ (and hence, $(\omega/\gamma)\,H_{\text{NSM-Q}}$) in the sector of $N_{\text{tot}}$ quasiparticles, provided that the transverse fields satisfy $d_Ad_B=-\omega\,g^{(1)}_{AB}$. The overall factor $\omega/\gamma$ slows down the effective dynamics, so that the time required to simulate the target grows linearly with the penalty strength, a general cost of perturbative gadgets~\cite{JordanFarhi2008,PhysRevA.91.012315}.

The self-energy is equally simple. The distance-zero processes flip a qubit $A$ into the sector $N_{\text{tot}}\pm1$ and back, and give
\begin{equation}
  \left.H_{\text{LR}}\,G_0\,H_{\text{RL}}\right|_{\text{diag}} = -\frac{1}{\gamma}\sum_A d_A^2\;\mathbb{L},
\end{equation}
which is the same for every logical state: it leaves the eigenstates unchanged, and we remove it from the energy levels with the uniform local field $\theta_A=(\omega N_\text{tot})^{-1}\sum_B d_B^2$, $\theta_{AB}=0$, so that the effective Hamiltonian in Eq. \eqref{eq:Heff_full} becomes
\begin{equation}
  \begin{split}
    H_{\text{eff}} = \frac{1}{\gamma}\Big[ \omega  & \Big(\sum_A h_A N_A + \sum_{A<B} g^{(2)}_{AB} N_A N_B \Big) \\
    &- \sum_{A<B} d_A d_B\left(X_A X_B + Y_A Y_B\right)\Big].
  \end{split}
  \label{eq:Heff_PC}
\end{equation}

Equation~\eqref{eq:Heff_PC} has the structure of Eqs.~\eqref{eq:H_NSMQ} and~\eqref{eq:H_FCI}, with one difference: the hopping is a rank-one matrix, $g^{(1)}_{AB}\to -d_Ad_B/\omega$. We therefore choose the transverse fields $\mathbf{d}$ that minimise
\begin{equation}
  \mathcal{L}(\mathbf{d}) = \sum_{A<B}\left(\omega \; g^{(1)}_{AB}+d_A d_B\right)^2 .
  \label{eq:loss_rank1}
\end{equation}
A generic $g^{(1)}$ has $N_q(N_q-1)/2$ independent entries, while the ansatz has only $N_q$ parameters, so $\mathcal{L}=0$ can be reached in general only for $N_q\le3$. Even for $N_q=3$, an exact solution requires $g^{(1)}_{12}g^{(1)}_{13}g^{(1)}_{23}<0$, because the product of the three rank-one couplings is $-(d_1d_2d_3)^2\le0$. For larger systems the encoding reproduces $g^{(1)}$ only approximately. We show below that, for the quasiparticle Hamiltonians considered here, the residual error still yields accurate ground-state fidelities.

\subsection{First Quantization Encoding (FQE)}
\label{sec:fqe}

The discrepancy between $g^{(1)}_{AB}$ and its rank-one approximation can be reduced systematically by changing the encoding of the logical space, which defines the FQE. In this case, we consider a Hilbert space $\mathcal{H}$ whose logical sector is spanned by first-quantized wavefunctions of the form
\begin{equation}
\ket{\Psi}=\sum_{a_1,\dots,a_{N_{\text{tot}}}}\Psi(a_1,\dots,a_{N_{\text{tot}}})\ket{a_1,\dots,a_{N_{\text{tot}}}},
\label{eq:Psi_FQ}
\end{equation}
where $a_{s}$ is the quasiparticle state of the $s$-th quasiparticle, encoded in a register $\mathcal{R}_{s}$ of $N_q$ qubits, so that $\mathcal{H}=\mathcal{R}_1\otimes\cdots\otimes\mathcal{R}_{N_{\text{tot}}}$. For example, if the first quasiparticle occupies the state $a$, then $a_1=a$ and the qubit $a$ of $\mathcal{R}_1$ is excited. The full Hilbert space $\mathcal{H}$ therefore has $N_G=N_{\text{tot}}N_q$ qubits and dimension $2^{N_G}$. 

In the logical subspace each register carries exactly one excitation, which identifies the state of the corresponding quasiparticle. Moreover, since the quasiparticles are hard-core bosons, two of them cannot occupy the same state, and $\Psi$ must be symmetric under their exchange. 

Each qubit is labelled by a register and a mode, $i\equiv(s,A)$, i.e.\ qubit $A$ of register $\mathcal{R}_s$, with occupation $N_i\equiv N^{(s)}_A$. The constraint reads
\begin{multline}
C=\sum_{s=1}^{N_{\text{tot}}}\Big[\Big(\sum_{A}N^{(s)}_A-1\Big)^2+\sum_{A<B}k_{AB}\,N^{(s)}_AN^{(s)}_B\Big]\\
+\sum_{s<s'}\sum_{A}N^{(s)}_AN^{(s')}_A .
\label{eq:C_FQ}
\end{multline}
The first term is the one-hot encoding of each register, and the last term enforces the hard-core condition between registers, with the same strength $\gamma$. The second term couples different states of the same register with dimensionless weights $k_{AB}$. It vanishes on the logical sector, where each register carries a single excitation, and only shifts the energy of the virtual states with several excitations in one register; its role in the effective couplings is discussed below. The transverse field is applied in the same way to every register,
\begin{equation}
H_X=\sum_{s=1}^{N_{\text{tot}}}\sum_{A}d_A X^{(s)}_A,
\label{eq:HX_FQ}
\end{equation}
with the same amplitudes $d_A$ in all registers.

Two logical states at Hamming distance two differ by moving one quasiparticle within a single register $\mathcal{R}_s$, from qubit $(s,A)$ to qubit $(s,B)$, through two virtual states of that register: the empty register $\ket{0}$ and the two-mode state $\ket{AB}$. Eq.~\eqref{eq:DeltaAB} then gives $\Delta_{(s,A)(s,B)}\equiv\Delta_{AB}$, the same for every register, with
\begin{equation}
\Delta_{AB}=\gamma\left(\frac{1}{E_{\ket{0}}-E_0}+\frac{1}{E_{\ket{AB}}-E_0}\right).
\label{eq:Delta_paths}
\end{equation}
The first term is fixed by the one-hot constraint, $E_{\ket{0}}-E_0=\gamma$, while the second is set by the intra-register couplings of Eq.~\eqref{eq:C_FQ}, $E_{\ket{AB}}-E_0=\gamma(1+k_{AB})$. Hence
\begin{equation}
\Delta_{AB}(k_{AB})=1+\frac{1}{1+k_{AB}},
\label{eq:Delta_k}
\end{equation}
which does not depend on the other quasiparticles, because neither virtual path involves the hard-core term. Focusing on the off-diagonal part of the effective Hamiltonian then reads 
\begin{multline}
H_{\text{eff}}\big|_{\text{off}}=-\frac{1}{2\gamma}\sum_{s=1}^{N_{\text{tot}}}\sum_{A<B}\bigg(d_Ad_B\,\Delta_{AB}(k_{AB})\\
\left(X^{(s)}_AX^{(s)}_B+Y^{(s)}_AY^{(s)}_B\right)\bigg).
\label{eq:Heff_FQ}
\end{multline}
For $k_{AB}=0$ both virtual paths cost $\gamma$, $\Delta_{AB}=2$, and Eq.~\eqref{eq:Heff_FQ} reduces to the rank-one coupling $-d_Ad_B/\gamma$ obtained with the particle-conservation constraint. For $k_{AB}\neq0$, instead, the weight of each pair can be tuned independently, which provides the additional freedom needed to go beyond the rank-one approximation. 

Now, starting from the optimal solution $\mathbf{d}^\star$ obtained from the rank-one loss from Eq.~\eqref{eq:loss_rank1}, we consider a new loss function
\begin{equation}
\mathcal{L}(\mathbf{k})=\sum_{A<B}\left|2 \omega \; g^{(1)}_{AB}+\Delta_{AB}(k_{AB})\,d^\star_Ad^\star_B\right|^2 .
\label{eq:loss_k}
\end{equation}
Each term depends on a single $k_{AB}$, so the minimum is found pair by pair in closed form,
\begin{equation}
\Delta_{AB}=-\frac{2 \omega \;g^{(1)}_{AB}}{d^\star_Ad^\star_B},\qquad k_{AB}=\frac{1}{\Delta_{AB}-1}-1 .
\label{eq:k_closed}
\end{equation}
Since Eq.~\eqref{eq:Delta_k} covers every real value except $\Delta_{AB}=1$, the global minimum $\mathcal{L}(\mathbf{k}^\star)=0$ is reached unless some pair requires exactly $\Delta_{AB}=1$. On the other hand, $k_{AB}$ also modifies the spectrum of the constraint, and it has to respect some conditions to preserve the validity of the Brillouin--Wigner expansion and the structure of the spectrum.

The first condition concerns the convergence of the expansion. At $k_{AB}=-1$ the state $\ket{AB}$ becomes degenerate with the logical sector and $\Delta_{AB}$ diverges. The expansion from Eq.~\eqref{eq:resolvent} is controlled only if this virtual state stays far from the logical sector on the scale of the drive, $\gamma\,|1+k_{AB}|\gg|d_A|,|d_B|$ for every pair (the convergence scale is set by the pair closest to the pole, $\min_{AB}(1+k_{AB})$), since the next-order corrections grow as $[d/\gamma(1+k_{AB})]^2$. 

The second condition concerns the order of the spectrum. Denoting by $N^{(s)}=\sum_A N^{(s)}_A$ the number of excitations in register $\mathcal{R}_{s}$, the one-hot and intra-register terms of Eq.~\eqref{eq:C_FQ} give the register energy
\begin{equation}
E^{(s)}=\gamma\Big[(N^{(s)}-1)^2+\sum_{A<B}k_{AB}N^{(s)}_AN^{(s)}_B\Big],
\label{eq:ES}
\end{equation}
while the hard-core term can only raise it.
For $k_{AB}<-1$ the state $\ket{AB}$ lies below the logical sector, which is no longer the ground space of $C$. A closed evolution prepared in the logical sector remains there as long as the sector is isolated, $\gamma|1+k_{AB}|\gg|d|$, but a forward anneal from the ground state of the driver ends outside it, and relaxation induced by a cold environment drives the system toward the unphysical states~\cite{Young2013}. For $k_{AB}>-1$, instead, all states with multiple excitations remain above the logical sector, since on a state with $N^{(s)}=n$ the sum in Eq.~\eqref{eq:ES} contains $n(n-1)/2$ terms, each larger than $-1$, so that $E^{(s)}>\gamma(n-1)(n-2)/2\ge\gamma$ for $n\ge3$. We therefore restrict to $k_{AB}>-1$, which corresponds to $\Delta_{AB}>1$, i.e.\ to targets with $-g^{(1)}_{AB}/(d^\star_Ad^\star_B)>1/2$. In particular, the value $k_{AB}=-2$, which gives $\Delta_{AB}=0$ and would switch off the pair, is not accessible, and vanishing target couplings must be approximated by a small residual value.

Finally, large positive values of $k_{AB}$ are stable but bring little, since $\Delta_{AB}\to1$ for $k_{AB}\to\infty$, while they enlarge the range of couplings that the device has to resolve. We therefore cap $k_{AB}\le k_{\max}$, with $k_{\max}=4$ ($\Delta_{AB}=1.2$). The matching is then exact for all pairs with $\Delta_{AB}\ge1.2$, while the remaining pairs, including those with the wrong sign, are set to $k_{\max}$ and keep a residual error, which is still smaller than in the rank-one fit.

The window $-2<k_{AB}<-1$, excluded above, is the only way the gadget can reverse the sign of an individual hopping. There Eq.~\eqref{eq:Delta_k} gives $\Delta_{AB}<0$: the path through $\ket{AB}$ has a negative energy denominator, and the generated hopping takes the sign of $+d_Ad_B$, so that a non-stoquastic $H_{\text{eff}}$ can be engineered. The price is paid in the structure of the spectrum. The states $\ket{AB}$ of the flipped pairs fall below the logical sector, which is no longer the ground space of $C$ but a band inside the spectrum of $H_G$, with unphysical states both above and below it. Any noise or relaxation that exchanges energy with the environment can then drive the system out of the logical space in both directions. Changing the sign of a coupling, and thereby making the effective Hamiltonian non-stoquastic, is thus paid for with a higher sensitivity to noise.

The diagonal terms of Eq.~\eqref{eq:H_qp} are added to the longitudinal part as in Eq.~\eqref{eq:HZ}. Since the occupation of mode $A$ is now $N_A=\sum_s N^{(s)}_A$, the one-body field $h_A$ is the same on qubit $A$ of every register, while the two-body coupling $g^{(2)}_{AB}$ acts only between different registers, $s\neq s'$, with $A\neq B$. Products within one register, $N^{(s)}_AN^{(s)}_B$, vanish on the logical sector and must be dropped, since an intra-register coupling would shift $k_{AB}$ and therefore $\Delta_{AB}$. The diagonal part of Eq.~\eqref{eq:HG_C} then reads
\begin{equation}
  \begin{split}
    H_{\text{diag}} = \frac{\omega}{\gamma}\Big[ &\sum_{s}\sum_A \left(h_A+\theta_A\right) N^{(s)}_A \\
    &+ \sum_{s<s'}\sum_{A\neq B}\left(g^{(2)}_{AB}+\theta_{AB}\right) N^{(s)}_A N^{(s')}_B \Big].
  \end{split}
  \label{eq:HZ_FQ}
\end{equation}

In this encoding the self-energy depends on the logical state. A flip $X^{(s)}_A$ of $\ket{a_1,\dots,a_{N_{\text{tot}}}}$ costs $\gamma$ if $A=a_{s}$, $\gamma(1+k_{a_{s}A})$ if $A\neq a_{s}$, and $\gamma(2+k_{a_{s}A})$ if, in addition, $A=a_{s'}$ is occupied by another quasiparticle. Summing over the registers,
\begin{equation}
  \left.H_{\text{LR}}\,G_0\,H_{\text{RL}}\right|_{\text{diag}}
  = \frac{1}{\gamma}\Big[\sum_{s} \sigma(a_{s}) + \sum_{s\neq s'}\varphi(a_{s},a_{s'})\Big],
  \label{eq:Sigma_FQ}
\end{equation}
with
\begin{align}
  \sigma(A) &= -d_A^2 - \sum_{B\neq A}\frac{d_B^2}{1+k_{AB}}, \label{eq:sigma_FQ}\\
  \varphi(A,B) &= \frac{d_B^2}{(1+k_{AB})(2+k_{AB})}. \label{eq:phi_FQ}
\end{align}
Here $\sigma$ is the self-energy of an isolated quasiparticle, and $\varphi$ is the correction due to the hard-core penalty. Equation~\eqref{eq:Sigma_FQ} is cancelled exactly at this order by
\begin{equation}
  \begin{split}
    \theta_A &= \frac{1}{\omega}\Big(d_A^2 + \sum_{B\neq A}\frac{d_B^2}{1+k_{AB}}\Big),\\
    \theta_{AB} &= -\frac{1}{\omega}\frac{d_A^2+d_B^2}{(1+k_{AB})(2+k_{AB})},
  \end{split}
  \label{eq:theta_FQ}
\end{equation}
which act on couplers already present in Eq.~\eqref{eq:HZ_FQ}. For $k_{AB}=0$, $\sigma$ is a constant, while the hard-core contribution depends on the state but has a one-body form, and is cancelled by $\theta_A=-(N_{\text{tot}}-1)\,d_A^2/(2\omega)$ alone, with $\theta_{AB}=0$.

With these choices the effective Hamiltonian in Eq. \eqref{eq:Heff_full} is, up to a constant,
\begin{equation}
  \begin{split}
    H_{\text{eff}} = \frac{1}{\gamma}\Big[ &\omega\Big(\sum_{s}\sum_A h_A N^{(s)}_A + \sum_{s<s'}\sum_{A\neq B} g^{(2)}_{AB} N^{(s)}_A N^{(s')}_B \Big) \\
    &- \frac12\sum_{s}\sum_{A<B} d_Ad_B\,\Delta_{AB}(k_{AB})\\
    &\times\left(X^{(s)}_AX^{(s)}_B+Y^{(s)}_AY^{(s)}_B\right)\Big],
  \end{split}
  \label{eq:Heff_FQ_full}
\end{equation}
and the residual error is of fourth order, $O(d^4/\gamma^3)$, since an odd number of flips cannot return to the logical sector. 

A last remark concerns the redundancy of the FQE. Since the quasiparticles are indistinguishable, the Hamiltonian is invariant under any permutation of the registers, and states such as $\ket{a_1,a_2}$ and $\ket{a_2,a_1}$ describe the same physical configuration. For two quasiparticles, the logical states split into symmetric and antisymmetric combinations, $\ket{a_1,a_2}\pm\ket{a_2,a_1}$, which are never mixed by the dynamics. Since the quasiparticles are hard-core bosons, only the symmetric sector is physical. This does not change the effective Hamiltonian, but it affects the preparation of the initial state: a single configuration $\ket{a_1,\dots,a_{N_{\text{tot}}}}$ is not symmetric, and the physical initial state is an entangled superposition of its $N_{\text{tot}}!$ permutations.

\begin{figure*}[tp]
    \centering
    \includegraphics[width=0.7\linewidth]{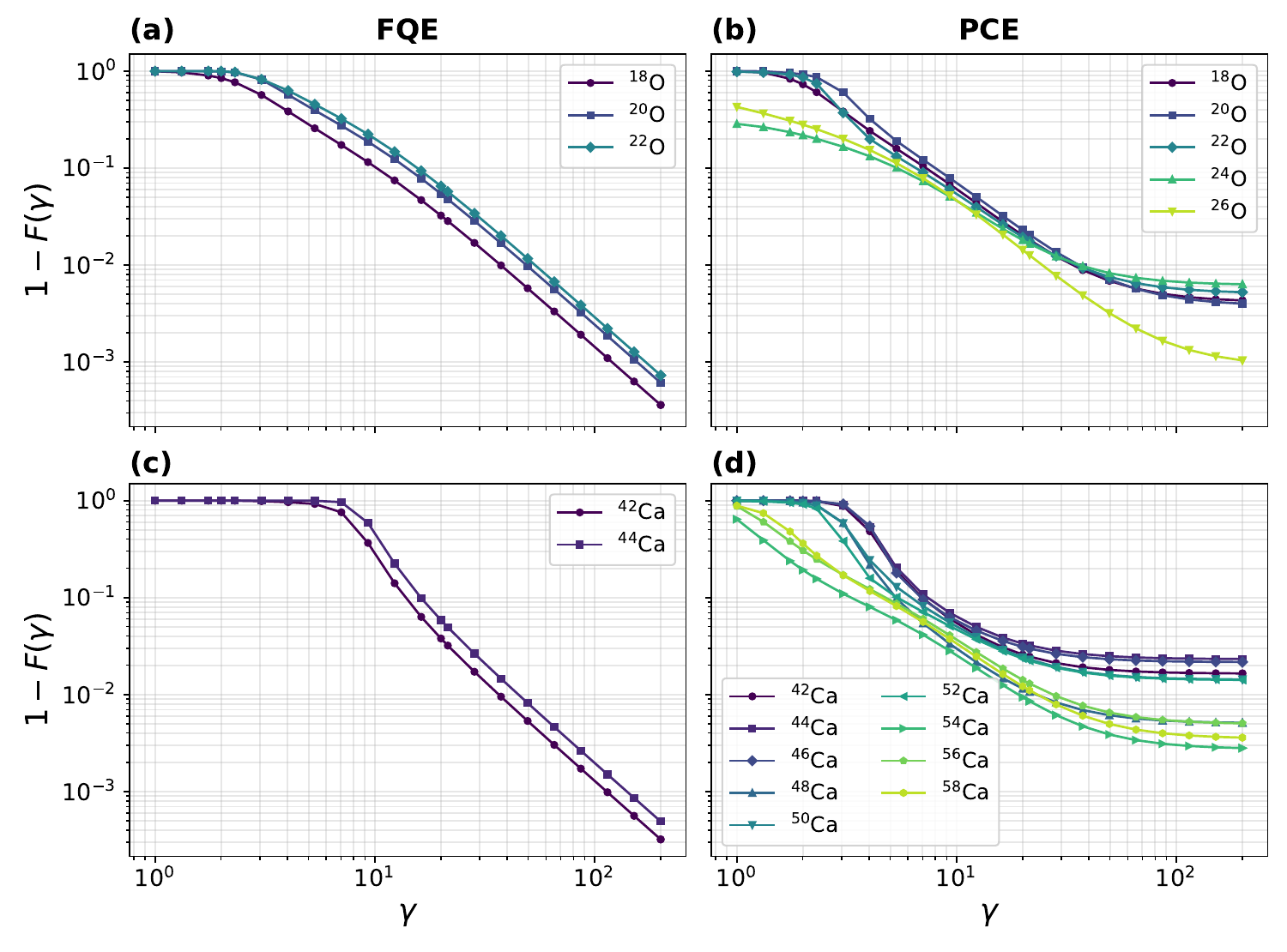}
    \caption{Ground-state infidelity $1-F(\gamma)$ versus the penalty strength $\gamma$ (in units of $\omega=1 \text{MeV}$) for the NSM quasiparticle Hamiltonians. (a), (c): FQE for the O and Ca isotopes with up to three and two quasiparticles, respectively. (b), (d): PCE for all non-trivial O ($N_G=6$) and Ca ($N_G=10$) isotopes. In the FQE, $1-F(\gamma)\sim\gamma^{-2}$ over the whole range; in the PCE, $1-F(\gamma)$ saturates at a $\gamma$-independent value set by the rank-one approximation of the hopping.}
    \label{fig:4}
\end{figure*}

\section{Benchmarking the Gadget Encodings}
\label{sec:results}

We now assess how well the gadget Hamiltonians reproduce the target quasiparticle Hamiltonians at the logical level, i.e., with arbitrary connectivity and qubit-dependent transverse fields. Hardware constraints are analyzed in Sec.~\ref{sec:embedding}. We address a static and a dynamic test. First, we quantify how close is the ground state $\ket{\Psi_G(\gamma)}$ of the gadget to the target ground state $\ket{\Psi_\text{gs}}$ using the fidelity $F(\gamma)=|\braket{\Psi_\text{gs}|\Psi_G(\gamma)}|^2$. Second, as a dynamic test, we check if this ground state can be prepared on the gadget itself by an annealing protocol, and how does the encoding affect the required time and the spectral gaps.

\subsection{NSM quasiparticle Hamiltonians}
\label{sec:results_nsm}

We consider the semi-magic oxygen ($sd$ shell) and calcium ($pf$ shell) isotopes, for which the quasiparticle description of Sec.~\ref{sec:nsm_qp_hamiltonian} is accurate \cite{Costa2026BW,COSTA2026140042}. The FQE requires $N_G=N_{\text{tot}}N_q$ qubits. To keep the numerical simulations tractable, we restrict it to at most three quasiparticles in oxygen, $^{18}$O ($N_G=6$), $^{20}$O ($12$), $^{22}$O ($18$), and two in calcium, $^{42}$Ca ($10$), $^{44}$Ca ($20$). The PCE uses $N_G=N_q$ qubits independently of $N_{\text{tot}}$, which allows us to treat all non-trivial isotopes, from $^{18}$O to $^{26}$O and from $^{42}$Ca to $^{58}$Ca.

Figure~\ref{fig:4} shows the infidelity $1-F(\gamma)$ for the two encodings (left and right columns) and for the two nuclei species (top and down rows). The results show three regimes:
\renewcommand{\theenumi}{\roman{enumi}} 
\begin{enumerate}
    \item \textit{Breakdown.}
Below a threshold $\gamma^\ast$, $1-F(\gamma)\simeq 1$: the penalty does not dominate the transverse fields, the perturbative expansion breaks down, and the gadget ground state lies outside $\mathcal{H}_L$. The threshold is larger for Ca ($\gamma^\ast\approx 8\,\omega$ in the FQE) than for O ($\gamma^\ast\approx 3\,\omega$).
\item \textit{Perturbative regime.}
Above $\gamma^\ast$, the infidelity decays as $1-F(\gamma)\sim\gamma^{-2}$, as expected from the leading corrections: the admixture of non-logical states into the ground state, with amplitude $O(d/\gamma)$, and the fourth-order terms of $H_{\text{eff}}$, whose relative weight is $O(d^2/\gamma^2)$. In the FQE this regime extends over the whole range, reaching $1-F(\gamma)<10^{-3}$ at $\gamma=200\,\omega$ for all isotopes.
\item \textit{Fitting floor.}
In the PCE the infidelity saturates at $1-F(\infty)$. All terms of $H_{\text{eff}}$ in the effective Hamiltonian from Eq.~\eqref{eq:Heff_PC} scale as $\omega^2/\gamma$, so increasing $\gamma$ only rescales its spectrum and leaves its eigenstates unchanged. For $\gamma\to\infty$ the gadget ground state therefore converges to the ground state of Eq.~\eqref{eq:Heff_PC}, which differs from the target only through the rank-one hopping $-d_Ad_B\approx \omega \; g^{(1)}_{AB}$. The plateau $1-F(\infty)$ is the infidelity caused by this approximation and vanishes if the loss in Eq.~\eqref{eq:loss_rank1} is zero. The plateaus remain low, $1-F(\infty)\in[10^{-3},10^{-2}]$ for O and $[3\times10^{-3},3\times10^{-2}]$ for Ca, with the smallest values for $^{26}$O, the particle-hole partner of $^{18}$O, and for the neutron-rich $^{54\text{--}58}$Ca. The FQE carries an analogous systematic error from the cutoff $k_{AB}\in\,]-1,k_{\max}]$, with $k_{\max}=4$, which is not visible in the range of $\gamma$ explored.
\end{enumerate}

The two encodings are therefore complementary: the FQE improves systematically with $\gamma$ over the explored range at a qubit cost linear in $N_{\text{tot}}$, while the PCE has a fixed qubit cost and an accuracy floor. However, for $\gamma\lesssim 30\,\omega$, the PCE yields the lowest infidelity, up to a factor $\sim6$ for $^{42}$Ca at $\gamma\approx10\,\omega$. This is the regime relevant for hardware: since $H_{\text{eff}}\propto \omega^2/\gamma$, increasing $\gamma$ compresses the logical spectrum relative to the largest programmed coupling, so the finite dynamic range and precision of the device bound the usable $\gamma$. 

Within the explored range, the accuracy does not degrade with the size of the problem: at $\gamma=200\,\omega$, the FQE stays below $7\times10^{-4}$ for every isotope, and the PCE floor varies by up to an order of magnitude between particle-hole partners such as $^{18}$O and $^{26}$O, which share the same $\dim(\mathcal{H}_L)$. No systematic size dependence is observed in the tested instances (see App.~\ref{app:scaling} for more details).

\subsection{LiH in DOCI}
\label{sec:results_lih}

The DOCI Hamiltonian poses a problem that does not arise in the nuclear case: its sign structure. The pair-hopping amplitudes of Eq.~\eqref{eq:H_DOCI} are exchange integrals, which are positive for real orbitals, so that $g^{(1)}_{AB}>0$ for every pair and every triangle of orbitals is frustrated, $g^{(1)}_{AB}g^{(1)}_{BC}g^{(1)}_{CA}>0$. As discussed in Sec.~\ref{sec:molecular_hamiltonian}, no sign gauge makes $H_\text{DOCI}$ stoquastic. The gadget, on the other hand, generates hoppings $-\tfrac12 d_Ad_B\Delta_{AB}$ with $\Delta_{AB}>0$ in the PCE and in the FQE with $k_{AB}>-1$. Their product around any triangle is non-positive, so $H_\text{eff}$ is stoquastic by construction, and no choice of transverse fields reproduces the full positive hopping matrix: the rank-one fit captures it only on a sign-consistent subset of pairs (see App.~\ref{app:stoq} for details). We therefore encode the sign-reversed Hamiltonian
\begin{equation}
\begin{split}
H_{\text{flip}} \equiv -H_\text{DOCI} = {} & -E_{\text{nucl}} + \sum_A h'_A N_A \\
& + \sum_{A<B} g'^{(1)}_{AB}\left(X_A X_B + Y_A Y_B\right) \\
& + \sum_{A<B} g'^{(2)}_{AB} N_A N_B ,
\end{split}
\label{eq:H_flip}
\end{equation}
with $g'^{(1,2)}_{AB}=-g^{(1,2)}_{AB}$ and $h'_A=-h_A$. The hopping of $H_\text{flip}$ is non-positive and is reproduced by transverse fields of equal sign.

The reversal leaves the eigenvectors unchanged but inverts the spectrum. The DOCI ground state becomes the highest eigenstate of $H_\text{flip}$ in $\mathcal{H}_L$. In a closed system this is not an obstacle in principle, since adiabatic evolution can follow the highest eigenstate of the band as well as the lowest one, provided it starts from the corresponding initial state. In an open system, however, relaxation towards low energies drives the population away from the target, leading to the ground state of $H_{\text{flip}}$ that corresponds with the highest energy state of $H_{\text{DOCI}}$. The sign flip therefore trades the sign problem for a higher sensitivity to noise, as for the FQE with $k_{AB}<-1$ (Sec.~\ref{sec:fqe}). In App.~\ref{app:stoq} we analyse the opposite choice, which keeps the target as the ground state of $H_G$ at the price of a stoquastic approximation of $H_\text{DOCI}$.

We consider LiH in the STO-3G basis, with $N_q=6$ spatial orbitals and $N_\text{tot}=2$ electron pairs, hence $\dim(\mathcal{H}_L)=15$, $N_G=6$ qubits in the PCE and $N_G=12$ in the FQE, along the bond length $R\in[0.5,4]$~\AA. For each $R$, $\ket{\Psi_G(\gamma,R)}$ is the highest state of the symmetric low-energy band of $H_G$ associated with the logical space, $H_G\ket{\Psi_G(\gamma,R)}=e_G(\gamma,R)\ket{\Psi_G(\gamma,R)}$ -with the specific counterterms added to remove the self-energy in both encodings-, and the corresponding molecular energy is
\begin{equation}
    E_G(\gamma,R) = -(\gamma/\omega)\, e_G(\gamma,R) + E_{\text{nucl}}(R) + N_\text{tot}\,\bar h(R),
    \label{eq:EG_lih}
\end{equation}
where the factor $\gamma$ undoes the $\omega/\gamma$ scaling of $H_\text{eff}$, the minus sign fixes the reversal of Eq.~\eqref{eq:H_flip}, and the constant restores $E_{\text{nucl}}$ and the mean orbital energy $\bar h=\sum_A h_A/N_q$ subtracted from the $h_A$ before encoding.

\begin{figure*}[tp]
    \centering
    \includegraphics[width=1\linewidth]{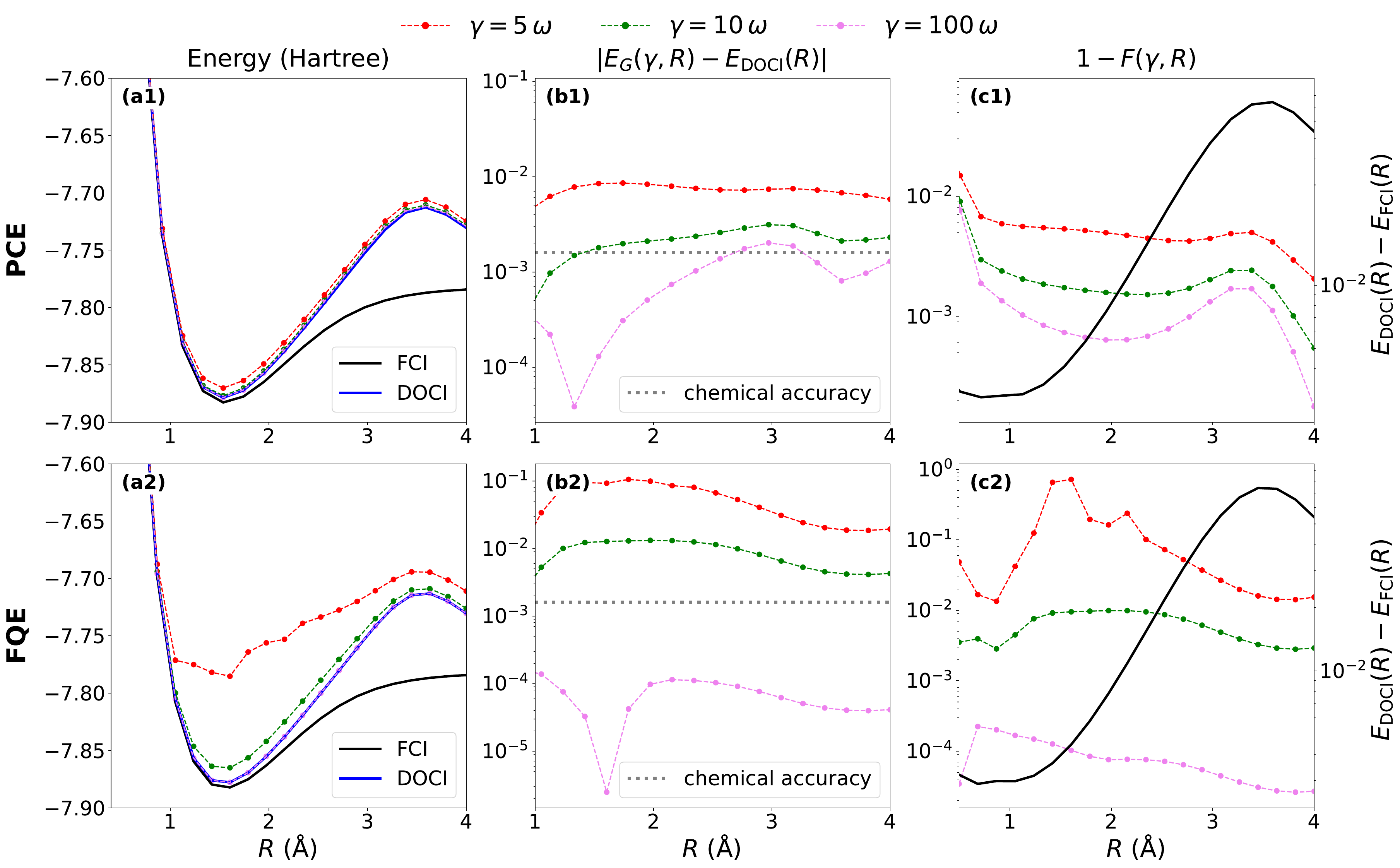}
    \caption{Gadget encoding of LiH (DOCI, STO-3G, canonical RHF orbitals) via the sign-reversed Hamiltonian $H_{\text{flip}}$, Eq.~\eqref{eq:H_flip}, along the bond length $R$. Top row: PCE ($N_G=6$); bottom row: FQE ($N_G=12$, $k_{\max}=4$). (a1,a2) Dissociation curve $E_G(\gamma,R)$ compared with $E_{\text{DOCI}}(R)$ and $E_{\text{FCI}}(R)$. (b1,b2) Energy error $|E_G(\gamma,R)-E_{\text{DOCI}}(R)|$; dotted line: chemical accuracy ($1.6$~mHa). Dips are sign changes of $E_G(\gamma,R)-E_{\text{DOCI}}(R)$. (c1,c2) Infidelity $1-F(\gamma,R)$ with respect to the DOCI ground state; black line, right axis: $E_{\text{DOCI}}(R)-E_{\text{FCI}}(R)$. Colours: $\gamma=5$, $10$, $100\,\omega$, with $\omega=1$~Ha.}
    \label{fig:lih_flip}
\end{figure*}

Figure~\ref{fig:lih_flip} compares $E_G(\gamma,R)$ with $E_\text{DOCI}(R)$ and $E_\text{FCI}(R)$ [panels (a)], and shows the energy error with respect to DOCI [(b)] and the infidelity with respect to the DOCI ground state [(c)], for $\gamma=5$, $10$ and $100\,\omega$. We refer chemical accuracy, $1.6$~mHa, to $E_\text{DOCI}(R)$, since the gadget targets $H_\text{DOCI}$ and not the full electronic problem.

The LiH results show the same three regimes found for the NSM Hamiltonians:
\begin{enumerate}
    \item \textit{Breakdown.}
Only the FQE enters it in the explored range. At $\gamma=5\,\omega$ its infidelity reaches $1-F(\gamma,R)\ge 0.5$ around equilibrium and the dissociation curve is lost [Fig.~\ref{fig:lih_flip}(a2,c2)]. The penalty $\gamma$ alone does not set the perturbative scale of the FQE: intra-register couplings with $k_{AB}<0$ lower the virtual states, $E_{\ket{AB}}-E_0=\gamma(1+k_{AB})$, so the effective separation is smaller than $\gamma$.
\item \textit{Perturbative regime.}
The PCE is in this regime already at $\gamma=5\,\omega$, where the dissociation curve keeps the shape of the DOCI curve and its equilibrium distance [Fig.~\ref{fig:lih_flip}(a1)]: between $5\,\omega$ and $10\,\omega$ its energy error drops from $5$--$8$~mHa to $0.5$--$3$~mHa, and it reaches its floor by $\gamma=100\,\omega$. For the FQE, from $\gamma=10\,\omega$ to $100\,\omega$ both $1-F(\gamma,R)$ and $|E_G(\gamma,R)-E_\text{DOCI}(R)|$ decrease by two orders of magnitude, compatible with $1-F(\gamma,R)\sim\gamma^{-2}$. At $\gamma=100\,\omega$ the FQE is within chemical accuracy along the whole dissociation curve, with $|E_G(\gamma,R)-E_\text{DOCI}(R)|\le0.14$~mHa and $1-F(\gamma,R)\le2.3\times10^{-4}$, and for $R\ge1$~\AA\ the infidelity decreases monotonically with $R$.
\item \textit{Fitting floor.}
In the PCE the errors at $\gamma=100\,\omega$ already coincide with their $\gamma\to\infty$ limit, set by the rank-one approximation of the hopping. The floor is low: near equilibrium $1-F(\gamma,R)=7\times10^{-4}$ and $|E_G(\gamma,R)-E_\text{DOCI}(R)|=0.17$~mHa, and the PCE leaves chemical accuracy only for $2.5\lesssim R\lesssim3.5$~\AA. The sharp dip of the energy error at $R\approx1.5$~\AA\ is a change of sign of $E_G(\gamma,R)-E_\text{DOCI}(R)$, not a point of exact encoding. The worst PCE fidelity, around $R\approx3.5$~\AA, lies close to the bond length where DOCI in canonical RHF orbitals is least accurate with respect to FCI [right axis of Fig.~\ref{fig:lih_flip}(c1)]. There the DOCI gap $\Delta_\text{gap}(R)=E^{(1)}_\text{DOCI}(R)-E_\text{DOCI}(R)$ is minimal, which amplifies the rank-one error in the fidelity. For the FQE, as in the NSM case, no saturation is visible in the explored range. However, the coupling cap at $k_{\max}=4$ leaves a finite mismatch between the effective Hamiltonian with fitted couplings and the target one, so $1-F(\gamma,R)$ must eventually saturate at a floor below the infidelities reached here.
\end{enumerate}

If we compare the two encoding methods, at $\gamma=10\,\omega$ we observed that the PCE is more accurate than the FQE, with energy errors of $0.5$--$3$~mHa against $5$--$10$~mHa, and infidelities of $\approx2\times10^{-3}$ against $3\times10^{-3}$--$10^{-2}$. This reproduces the behaviour of the NSM Hamiltonians for $\gamma\lesssim30\,\omega$. At $\gamma=100\,\omega$, instead, the FQE improves on the PCE by about one order of magnitude in both $1-F(\gamma,R)$ and $|E_G(\gamma,R)-E_\text{DOCI}(R)|$. Since the device bounds the usable $\gamma$ (Sec.~\ref{sec:results_nsm}), the PCE, which also needs half the qubits, is the natural choice on current hardware, while the FQE is the route to systematically improvable accuracy.

Finally, the encoding error should be compared with the error of the model itself. DOCI in canonical RHF orbitals deviates from FCI by $5$~mHa at equilibrium and by up to $80$~mHa at $R\approx3.5$~\AA, where it also produces the unphysical maximum of the dissociation curve visible in Fig.~\ref{fig:lih_flip}(a). The canonical RHF orbitals, optimised for a single closed-shell determinant, are inadequate for the stretched bond, and the maximum of $E_\text{DOCI}(R)-E_\text{FCI}(R)$ marks where they fail most. At $\gamma=100\,\omega$ the encoding error is smaller than this model error at every $R$, in both PCE and FQE. With canonical orbitals, the accuracy of the gadget is thus limited by the model and not by the encoding.

\subsection{Adiabatic state preparation}
\label{sec:qa}

\begin{figure*}[tp]
    \centering
    \includegraphics[width=1\linewidth]{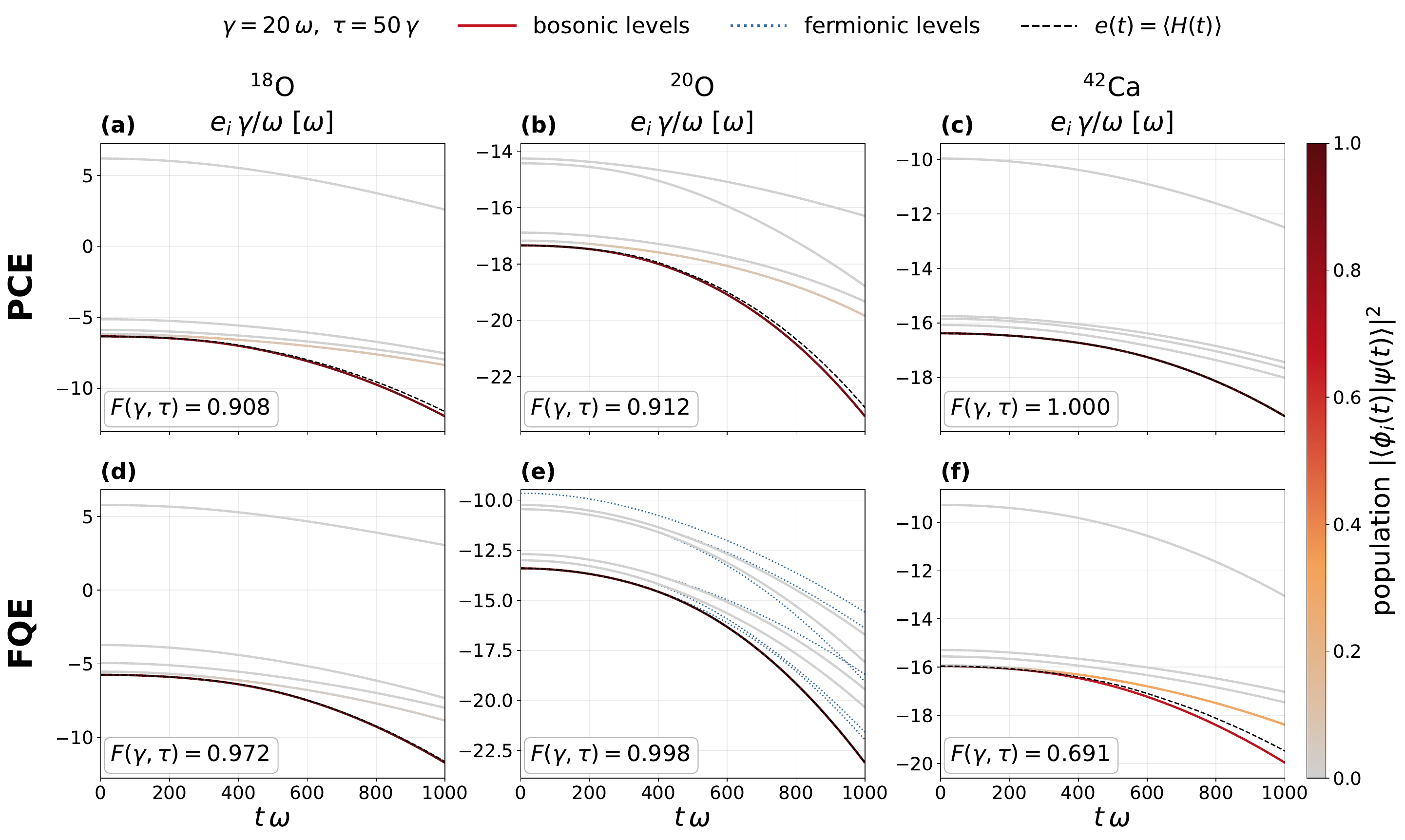}
    \caption{Adiabatic preparation of the gadget ground state with the protocol of Eq.~\eqref{eq:H_t}, $\gamma=20\,\omega$ and $\tau=50\,\gamma/\omega^2$, for $^{18}$O, $^{20}$O and $^{42}$Ca. Top row (a--c): PCE; bottom row (d--f): FQE. Lines: instantaneous eigenvalues $e_i(t)$ of $H(t)$, in units of $\omega^2/\gamma$, coloured by the population $|\braket{\phi_i(t)|\psi(t)}|^2$; solid: bosonic levels, dotted: fermionic levels (FQE with $N_\text{tot}=2$); dashed black: mean energy $e(t)=\braket{\psi(t)|H(t)|\psi(t)}$. The final fidelity $F(\gamma,\tau)$ is reported in each panel.}
    \label{fig:qa}
\end{figure*}

A high fidelity is necessary but not sufficient as the encoded ground state must also be reachable on the device. Since the gadget is an Ising Hamiltonian with transverse fields, it can be prepared by an annealing protocol, which is the native operation of the hardware. 
We keep the longitudinal part fixed and switch on the transverse fields,
\begin{equation}
    H(t) = H_Z + B(t)\,H_X, \quad B(t)=\frac{t}{\tau}, \quad t\in[0,\tau],
    \label{eq:H_t}
\end{equation}
starting from the ground state of $H_Z$, the lowest-energy quasiparticle configuration in $\mathcal{H}_L$, and ending in $H(\tau)=H_G$: the first half of a reverse-annealing cycle~\cite{Perdomo2011,Ohkuwa2018,Yamashiro2019}. The protocol differs from Ref.~\cite{costa2024quantum} in two ways. First, in the previous protocol the annealing interpolates between a longitudinal driver, designed to keep a large gap, and the full NSM Hamiltonian. Here, the driver is the longitudinal part $H_Z$ of the gadget Hamiltonian, which is never switched off. Second, the hopping is generated by the transverse field itself at second order, so the effective schedule in $\mathcal{H}_L$ is quadratic, $(B(t)\,d)^2/\gamma$, rather than linear. Moreover, since the logical spectrum scales as $\omega^2/\gamma$, in this approach the adiabatic time grows linearly with $\gamma$, and we set $\tau=50\,\gamma/\omega^2$.

The initial state is simple to prepare in the PCE: a single bitstring. In the FQE with $N_\text{tot}\ge2$, instead, the ground state of $H_Z$ is degenerate among the states obtained by permuting the registers. $H(t)$ commutes with these permutations, so the symmetric (bosonic) and antisymmetric (fermionic) sectors evolve independently, and only the bosonic sector describes the hardcore-boson quasiparticles. The initial state must then be the symmetric superposition of the permuted bitstrings, an entangled state. It can be prepared by a standard forward annealing from a uniform transverse field to the longitudinal part of the gadget, which preserves the permutation symmetry and ends in the symmetric ground state. Here we focus on the reverse annealing of Eq.~\eqref{eq:H_t}, and the FQE simulations start from the already prepared symmetric ground state of $H_Z$.
As a figure of merit we use the fidelity between the final state and the gadget ground state, $F(\gamma,\tau)=|\braket{\Psi_{G}(\gamma)|\psi(\tau)}|^2$.

Figure~\ref{fig:qa} shows the instantaneous spectrum of $H(t)$, $H(t)\ket{\phi_i(t)}=e_i(t)\ket{\phi_i(t)}$, for $^{18}$O, $^{20}$O and $^{42}$Ca at $\gamma=20\,\omega$, together with the populations $|\braket{\phi_i(t)|\psi(t)}|^2$ and the mean energy $e(t)=\braket{\psi(t)|H(t)|\psi(t)}$. Energies are shown in units of $\omega^2/\gamma$, so that the logical spectra at different $\gamma$ are directly comparable.

Three of the six runs reach $F(\gamma,\tau)>0.95$: the FQE for both oxygen isotopes ($F(\gamma,\tau)\geq0.97$; in $^{20}$O the fermionic levels cross excited bosonic ones without population transfer, Fig.~\ref{fig:qa}(e)) and the PCE for $^{42}$Ca ($1-F(\gamma,\tau)\simeq10^{-5}$). In the other three, about $9\%$ (PCE, $^{18}$O and $^{20}$O, $F(\gamma,\tau)\approx0.91$) and $30\%$ (FQE, $^{42}$Ca, $F(\gamma,\tau)=0.691$) of the population ends in the first excited level. 

In contrast with Ref.~\cite{costa2024quantum}, where the minimum gap is reached at the end of the annealing, here the smallest gaps between the populated levels occur in the first half of the evolution, $t/\tau\lesssim0.5$. We attribute this change to the different structure of the protocol: a reverse annealing from $H_Z$, which already contains the full diagonal part of the target and the self-energy counter-terms, while the hopping, generated at second order, is switched on with the effective quadratic schedule $B(t)^2=(t/\tau)^2$. 

More generally, the protocol exposes a trade-off in the choice of $\gamma$. A larger $\gamma$ improves the static accuracy of the encoding, but compresses the logical spectrum as $\omega^2/\gamma$, which lengthens the required annealing time linearly in $\gamma$ and tightens the precision required of the device.

\section{Embedding in state-of-the-art annealers}
\label{sec:embedding}

So far we observed that the main sources of complexities for this implementation are related to the perturbation regime, with a large energy resolution window $[\frac{\omega^2}{\gamma},\gamma]$, a rescaled time evolution in $\gamma/\omega$, and a lack of precision in the reconstruction of the correct couplings. However, when we deal with quantum annealers, we may encounter two other important challenges that may increase the limits of this implementation: a native fixed topology of the device, and the lack of tunable local transverse fields. In this section we show how it is possible to take into account these limitations in the framework of the gadget ising embedding and we study the regimes and the limitations under these conditions.

\subsection{Fixed connectivity and minor embedding}
\label{sec:minor_embedding}

In both encodings the $N_iN_j$ interactions of $H_G$ are dense: the PCE is fully connected, while in the FQE full connectivity is restricted to each register $\mathcal{R}_s$, with inter-register couplings set by $g^{(2)}_{ij}$ and the hardcore boson constrain. No commercial annealer can represent this connectivity natively, so we adopt the minor-embedding strategy~\cite{Choi2008,Cai2014} to reduce the connectivity of $H_G$. For computational feasibility, we carry out the embedding analysis for the PCE. The method applies equally to the FQE, which requires more qubits but has a sparser connectivity. In the PCE each gadget qubit encodes one quasiparticle mode, $i\equiv A$ (Sec.~\ref{sec:pce}), and from here on gadget qubits are labelled by the mode indices $A,B$. We represent each gadget qubit $q^{G}_{A}$ using $L_{A}$ physical qubits $q^{\text{Ph}}_{A,r}$, $r=1,\dots,L_A$, connected by a strong ferromagnetic coupling $J_F$. Each chain is described by
\begin{multline}
    H_A^{\text{ch}} = -J_F\sum_{r=1}^{L_{A}-1} Z_{A,r}Z_{A,r+1} \\
    + \sum_{r=1}^{L_{A}}\left(\Gamma_{A,r}X_{A,r} + \Xi_{A,r}N_{A,r}\right),
    \label{eq:H_chain}
\end{multline}
where $\Gamma_{A,r}$ and $\Xi_{A,r}$ are the programmable transverse and longitudinal fields acting on the qubit $q^{\text{Ph}}_{A,r}$ of the chain $A$.
For $J_F\gg|\Gamma_{A,r}|,|\Xi_{A,r}|$, the low-energy space of the chain is spanned by the two ferromagnetic states $\ket{0\cdots0}$ and $\ket{1\cdots1}$, which encode the states $\ket{0}$ and $\ket{1}$ of the gadget qubit, with $N_{A}=N_{A,r}$ for every $r$. The longitudinal fields simply add up, whereas the transverse fields connect the two ferromagnetic states only at order $L_{A}$, by flipping all the physical qubits one at a time through intermediate states with one or more broken bonds, each costing $2J_F$. Summing over the $L_{A}!$ flip orders (see App.~\ref{app:chain}), the chain acts on the gadget qubit as
\begin{equation}
    \begin{split}
    H_A^{\text{eff}} &= \tilde d_A X_A + \tilde h_A N_A, \\
    \tilde d_A &= (-1)^{L_A-1}\frac{\prod_{r}\Gamma_{A,r}}{J_F^{\,L_A-1}}, \qquad
    \tilde h_A = \sum_{r}\Xi_{A,r},
    \end{split}
    \label{eq:chain_eff}
\end{equation}
up to a constant, where $X_A$ flips the whole chain. The transverse field is therefore renormalized by the factor $(\Gamma/J_F)^{L_A-1}$ and its sign alternates with the chain length. The same result holds for chains with the shape of a tree and bond-dependent couplings $J_e$, with $J_F^{L_A-1}\to\prod_e J_e$. To reproduce $H_G$ we set $\tilde d_A=d_A$ and $\tilde h_A=\Xi_A$, the coefficient of $N_A$ in $H_Z$, and distribute both uniformly along the chain,
\begin{equation}
    \begin{split}
    |\Gamma_{A,r}| &= \left(J_F^{\,L_A-1}|d_A|\right)^{1/L_A}, \ 
    \Xi_{A,r} = \frac{\Xi_A}{L_A}, \\
    \prod_r\operatorname{sgn}\Gamma_{A,r} &= (-1)^{L_A-1}\operatorname{sgn}(d_A).
    \end{split}
    \label{eq:chain_fields}
\end{equation}
The rescaled fields keep the effective Hamiltonian at the scale $\omega^2/\gamma$, but the perturbative condition along the chain, $|\Gamma_{A,r}|/J_F=\epsilon\ll1$, requires $J_F=|d_A|\,\epsilon^{-L_A}$: the dynamic range of the device grows exponentially with the chain length.

We rewrite $H_Z=\gamma C+H_{\text{diag}}$, Eq.~\eqref{eq:HG_C}, in the PCE by combining the linear and quadratic terms of $C$, Eq.~\eqref{eq:C_PC}, and of $H_{\text{diag}}$, Eq.~\eqref{eq:HZ}, with $\theta_A=\theta_{AB}=0$, since in the PCE the self-energy is a global shift,
\begin{equation}
    \begin{split}
    H_Z &= \sum_A \Xi_A N_A + \sum_{A<B} \Lambda_{AB}\, N_AN_B + \text{const}, \\
    \Xi_A &= \gamma\left(1-2N_\text{tot}\right) + \frac{\omega\, h_A}{\gamma}, \ 
    \Lambda_{AB} = 2\gamma + \frac{\omega\, g^{(2)}_{AB}}{\gamma}.
    \end{split}
    \label{eq:HZ_PC_emb}
\end{equation}
Each gadget qubit $A$ is replaced by a chain with bonds $\mathcal{E}_A$, and each coupling $\Lambda_{AB}$ by the set $\mathcal{E}_{AB}$ of physical couplers between the chains $A$ and $B$ provided by the embedding. The physical Hamiltonian is
\begin{equation}
    \begin{split}
    H_G^{\text{Ph}} = &-J_F\sum_A\sum_{(r,r')\in\mathcal{E}_A}Z_{A,r}Z_{A,r'} \\
    &+\sum_A\sum_{r=1}^{L_A}\Big(\Gamma_{A,r}X_{A,r}+\frac{\Xi_A}{L_A}N_{A,r}\Big) \\
    &+\sum_{A<B}\frac{\Lambda_{AB}}{|\mathcal{E}_{AB}|}\sum_{(r,s)\in\mathcal{E}_{AB}}N_{A,r}N_{B,s},
    \end{split}
    \label{eq:HG_PC_emb}
\end{equation}
acting on $N_\text{Ph}=\sum_A L_A$ physical qubits, with $\Gamma_{A,r}$ given by Eq.~\eqref{eq:chain_fields}. In the frozen-chain manifold, equation~\eqref{eq:HG_PC_emb} reduces to the PCE gadget $H_G$. During the virtual flips of a chain, the couplers to the occupied chains shift the energy of the intermediate states by up to $\sim N_\text{tot}\gamma$, so that the hierarchy of scales becomes $J_F\gg N_\text{tot}\gamma$, $\gamma\gg|d_A|$. To quantify the effect of the minor embedding, we compute by exact diagonalization the ground state $\ket{\Psi_G(J_F,\gamma)}$ of $H_G^{\text{Ph}}$, Eq.~\eqref{eq:HG_PC_emb}, for different values of $J_F$ and $\gamma$. Since each chain has only two frozen states, every logical configuration of the PCE corresponds to exactly one physical configuration, in which all the qubits of the occupied chains are set to $1$. This one-to-one correspondence defines a mapping $\mathbb{E}$ that encodes the target ground state $\ket{\Psi_{\text{gs}}}$ of the quasiparticle Hamiltonian in the frozen-chain subspace, and we compute the fidelity
\begin{equation}
    F(J_F,\gamma)=\left|\braket{\Psi_{\text{gs}}|\mathbb{E}^\dagger|\Psi_G(J_F,\gamma)}\right|^2,
    \label{eq:F_emb}
\end{equation}
so that any weight on broken chains or outside the logical space counts as infidelity.

\begin{figure*}
    \centering
    \includegraphics[width=\linewidth]{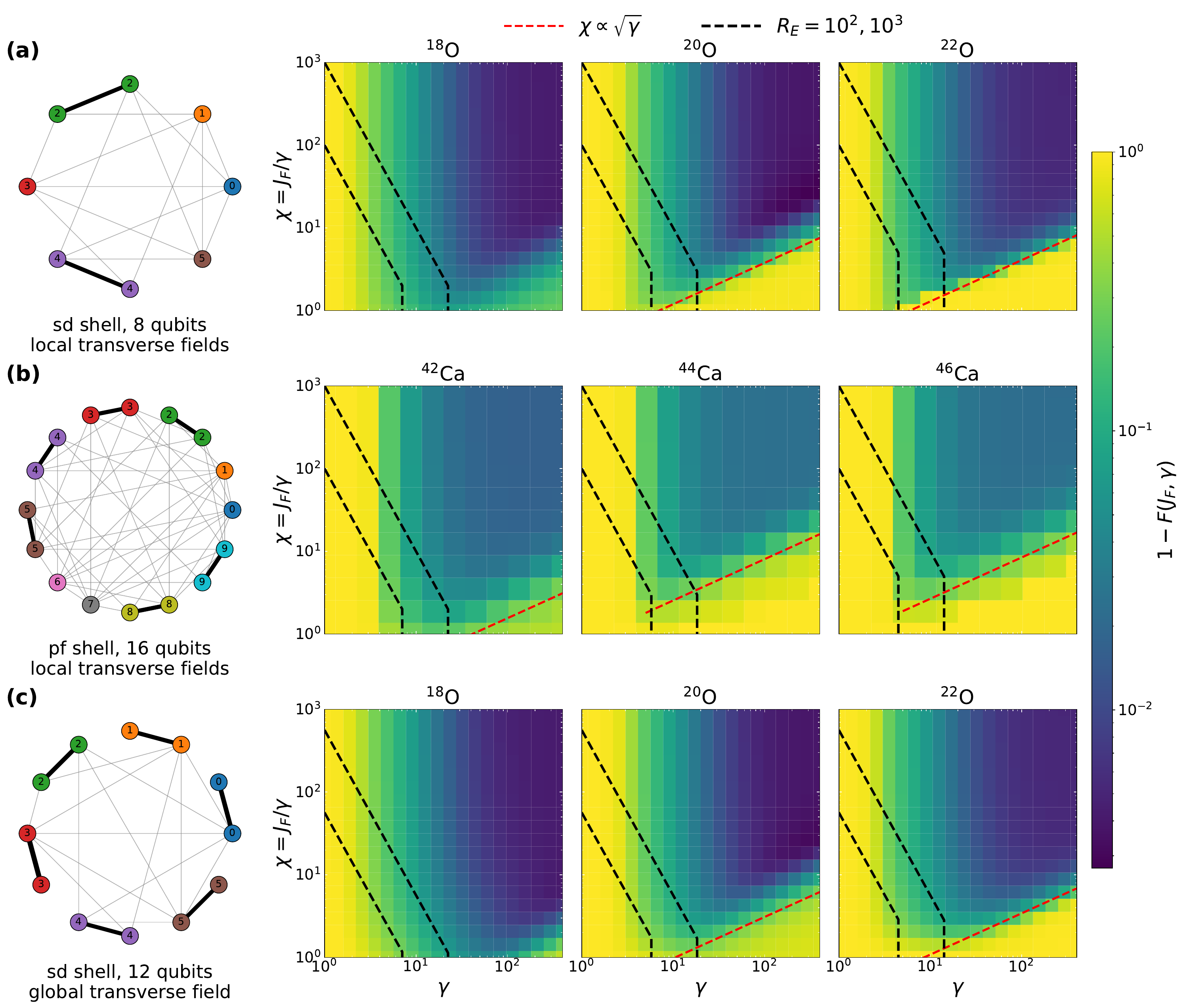}
    \caption{Left: Pegasus minor embeddings of the PCE gadget for (a) the $sd$ shell with local transverse fields (8 physical qubits), (b) the $pf$ shell with local transverse fields (16 physical qubits), and (c) the $sd$ shell with a global transverse field (12 physical qubits). Nodes are physical qubits labelled by their gadget qubit $A$; thick edges are the ferromagnetic chain bonds $J_F$, thin edges the inter-chain couplers. In (c) all gadget qubits are chains of $L=2$ physical qubits driven by the same field $\Gamma=\sqrt{\omega J_F}$, with chain bonds $J_{F,A}=f_AJ_F$ from Eq.~\eqref{eq:fA_global}. Right: ground-state infidelity $1-F(J_F,\gamma)$, Eq.~\eqref{eq:F_emb}, versus the penalty $\gamma$ (in units of $\omega=1$ MeV) and the ratio $\chi=J_F/\gamma$, for $^{18,20,22}$O [(a), (c)] and $^{42,44,46}$Ca [(b)]. For $\chi\gtrsim10^2$ the unembedded result $1-F(\gamma)$ of Fig.~\ref{fig:4}(b),(d) is recovered. Red dashed lines: breakdown boundary $\chi\propto\sqrt{\gamma}$, Eq.~\eqref{eq:gamma_star}. Black dashed lines: curves of constant $R_E=\chi\gamma^2/\omega^2$, Eq.~\eqref{eq:R_range}, for $R_E=10^2$ and $10^3$.}
    \label{fig:embedding_scan}
\end{figure*}

We embed the PCE gadget into the Pegasus topology of D-Wave annealers~\cite{Boothby2020} with the \texttt{minorminer} heuristic~\cite{Cai2014}, obtaining $N_\text{Ph}=8$ (O) and $N_\text{Ph}=16$ (Ca) physical qubits, with two and six chains of $L_A=2$ qubits, respectively (left panels of Fig.~\ref{fig:embedding_scan}). We use a single embedding at fixed seed since, by Eq.~\eqref{eq:chain_eff}, at leading order the chains enter the effective Hamiltonian only through their lengths $L_A$. As $H_G$ has the same structure for the NSM and the DOCI Hamiltonians, we restrict the analysis to the O and Ca isotopes.

Figures~\ref{fig:embedding_scan}(a),(b) show the infidelity in the plane $(\gamma,\chi)$, with $\chi=J_F/\gamma$, and reveal three regimes:
\renewcommand{\theenumi}{\roman{enumi}}
\begin{enumerate}
    \item \textit{Gadget breakdown}: for $\gamma\lesssim4\,\omega$, $1-F(J_F,\gamma)\simeq1$ for every $\chi$: the gadget condition $\gamma\gg|d_A|$ itself fails [Fig.~\ref{fig:4}(b),(d)], and no chain strength can restore the fidelity.
    \item \textit{Transparent embedding}: for $\chi\gtrsim10^2$, $F(J_F,\gamma)$ depends on $\gamma$ only and reproduces the unembedded $F(\gamma)$ of Fig.~\ref{fig:4}(b),(d), including the plateau set by the rank-one approximation of the hopping, $1-F(\gamma)\simeq4$--$5\times10^{-3}$ for O and $1.7$--$2.3\times10^{-2}$ for Ca. For sufficiently stiff chains the minor embedding is therefore transparent, and the accuracy is limited by the gadget encoding alone.
    \item \textit{Inter-chain collapse}: at finite $\chi$, the fidelity first improves with $\gamma$ and then collapses beyond a critical value $\gamma^*(\chi)$. The collapse is caused by the inter-chain terms of Eq.~\eqref{eq:HG_PC_emb}: being of order $\gamma$ and acting on specific qubits of each chain, they change the energies of the virtual chain-breaking processes in a configuration-dependent way. 
\end{enumerate}

For the last regime, and as shown in App.~\ref{app:asym}, the critical value $\gamma^*$ is
\begin{equation}
    \gamma^*\sim\frac{\Delta_G}{|d_A|}\,\chi^2 \, \omega
    \quad\Longleftrightarrow\quad
    \gamma^*\propto J_F^{2/3} \omega^{1/3},
    \label{eq:gamma_star}
\end{equation}
where $\Delta_G$ is the gap of the rescaled effective Hamiltonian $(\gamma/\omega) H_{\text{eff}}$ of Eq.~\eqref{eq:Heff_full}. 
The chain strength must therefore grow faster than the penalty, $J_F\propto\gamma^{3/2}  \omega^{-1/2}$, and the stable region is wider for effective Hamiltonians with a larger gap $\Delta_G$. Indeed, $\Delta_G\simeq3.7$--$4.2$~MeV for the O isotopes and $\Delta_G\simeq1.3$--$1.4$~MeV for the Ca isotopes, consistent with the wider stable region of the former. 

The gap is not the only factor, however, that produces the previous effect: the prefactor in Eq.~\eqref{eq:gamma_star} also depends on the occupation of the logical states and on how the couplers are distributed along the chains, so that $\gamma^*$ can differ between isotopes with similar $\Delta_G$. For $N_\text{tot}\ge2$ ($^{20,22}$O, $^{44,46}$Ca) the collapse is sharp and follows Eq.~\eqref{eq:gamma_star} (red dashed lines). For a single pair ($^{18}$O, $^{42}$Ca), instead, $\gamma^*$ is barely visible in the scanned range and the embedding remains stable down to $\chi\sim1$. Since $^{42,44,46}$Ca have almost the same gap, this robustness is not a gap effect, and we relate it to the condition $J_F\gg N_\text{tot}\gamma$: the more chains are occupied, the more inter-chain terms contribute to the cost of a chain breaking.

Finally, Fig.~\ref{fig:embedding_scan} identifies the operating region of the embedding under the assumed range of hardware coefficients. On an annealer all the terms of $H_G^{\text{Ph}}$ must fit in the same programmable range. Its maximum is set by the largest coefficient of $H_G^{\text{Ph}}$: the chain coupling $J_F=\chi\gamma$, the penalty couplers $\Lambda_{AB}\simeq2\gamma$, or the longitudinal fields $\Xi_A\simeq\gamma(1-2N_\text{tot})$, which grow with the number of pairs. The smallest energy that the gadget must resolve is instead the logical scale $\omega^2/\gamma$. We therefore define the ratio
\begin{equation}
    R_E=\frac{\gamma^2}{\omega^2}\max\Big(\chi,\,2,\,\max_A\frac{|\Xi_A|}{\gamma}\Big),
    \label{eq:R_range}
\end{equation}
which must stay below the inverse of the relative precision with which the device sets its couplings; this is a necessary condition for the logical energies to be resolved, not a guarantee of $F(J_F,\gamma)$. Stronger chains improve the embedding but increase $R_E$, compressing the logical energies, the same trade-off found for other minor-embedded problems~\cite{Choi2008,Venturelli2015}. 

In the plane of Fig.~\ref{fig:embedding_scan}, in the region where $J_F$ is the largest coefficient, the curves of constant $R_E$ are the lines $\chi=R_E\,\omega^2/\gamma^2$ (black dashed), and only the region below them is accessible. The maximum fidelity along each curve lies close to the corner between the breakdown of the gadget at small $\gamma$ and the breakdown of the chains at $\gamma^*$. For $R_E=10^2$ we obtain $F(J_F,\gamma)\simeq 0.72$--$0.84$ for the O isotopes, $F(J_F,\gamma)\simeq0.73$ for $^{42}$Ca and $F(J_F,\gamma)\simeq0.69$ for $^{44}$Ca, but only $F(J_F,\gamma)\lesssim0.13$ for $^{46}$Ca. For $R_E=10^3$, we obtain $F(J_F,\gamma)\simeq 0.94$--$0.96$ for the O isotopes and $F(J_F,\gamma)\simeq0.91$--$0.94$ for the Ca isotopes. The asymptotic region of large $\gamma$ and $\chi$ requires a much larger $R_E$. Equation~\eqref{eq:chain_fields}, however, requires qubit-dependent transverse fields, whereas current D-Wave devices provide only a global drive; Sec.~\ref{sec:global_field} shows how to lift this requirement.

\subsection{Global transverse field for stoquastic quasiparticle Hamiltonians}
\label{sec:global_field}

Commercial annealers do not provide independently tunable transverse fields: the driver acts with the same amplitude $\Gamma\ge0$ on every physical qubit. The gadget, instead, requires a qubit-dependent drive $d_AX_A$. The chain construction of Sec.~\ref{sec:minor_embedding} also solves this problem. By Eq.~\eqref{eq:chain_eff}, the effective drive of a chain is fixed by its bonds as much as by its fields, so the dependence on $A$ can be moved from the transverse fields to the ferromagnetic couplers, which are individually programmable. We therefore represent every gadget qubit by a chain of $L_A\ge2$ physical qubits with a chain-specific bond $J_{F,A}$, and set $\Gamma_{A,r}=\Gamma$ in Eq.~\eqref{eq:HG_PC_emb}. Equation~\eqref{eq:chain_eff}, with $J_F\to J_{F,A}$, gives
\begin{equation}
    \tilde{d}_A=(-1)^{L_A-1}\,\frac{\Gamma^{L_A}}{J_{F,A}^{\,L_A-1}}.
    \label{eq:d_global}
\end{equation}
A single-qubit chain, $L_A=1$, gives $\tilde d_A=\Gamma$ for every $A$ and cannot be tuned, which is why every gadget qubit, including those that need no chain for connectivity, must be embedded into a chain with $L_A\ge2$.

The global field fixes the sign of every $\tilde d_A$ to $(-1)^{L_A-1}$. This does not restrict stoquastic Hamiltonians, which the gadget reproduces regardless of the signs of the drives. For the NSM, a suitable choice of the phases of the quasiparticle states makes all the couplings $g^{(1)}_{AB}$ negative, and drives of equal sign, $\tilde d_A=-|d_A|$ for $L_A=2$, reproduce them. In this gauge all ground-state amplitudes are non-negative and $F(J_F,\gamma)$ is computed directly. Non-stoquastic Hamiltonians require instead a sign flip of the effective couplings, which is set by the position of the virtual states relative to the logical space (Sec.~\ref{sec:results_lih}), not by the sign of $\Gamma$.

The chain lengths matter instead for the magnitudes. Since $\Gamma$ is common to all chains, we take a uniform length $L_A=L$, write $J_{F,A}=f_AJ_F$, and rescale the field as $\Gamma=\omega^{1/L}J_F^{(L-1)/L}$, so that Eq.~\eqref{eq:d_global} becomes
\begin{equation}
    \tilde{d}_A=(-1)^{L-1}\,\frac{\omega}{f_A^{\,L-1}},
    \qquad
    f_A=\left(\frac{\omega}{|d_A|}\right)^{1/(L-1)},
    \label{eq:fA_global}
\end{equation}
independent of $J_F$. The perturbative parameter of the chains is $\Gamma/J_{F,A}=(\omega/J_F)^{1/L}/f_A$. Requiring it to be $\simeq\epsilon$, so that $\omega^2/\gamma$ remains the only scale of the dynamics, fixes the energy scale of the ferromagnetic couplings, as in the local-field case, to $J_F\simeq\omega/(f_A\epsilon)^{L}$.
We use the minimal length $L=2$, obtained from the local-field embeddings of Fig.~\ref{fig:embedding_scan}(a) by extending every single-qubit chain to a free neighbouring qubit of the Pegasus graph. The number of physical qubits grows from $N_\text{Ph}=8$ to $N_\text{Ph}=12$ for the sd shell and from $N_\text{Ph}=16$ to $N_\text{Ph}=20$ for the pf shell. 

Figure~\ref{fig:embedding_scan}(c) shows the infidelity $1-F(J_F,\gamma)$ of the global-field embedding for $^{18,20,22}$O, in the same $(\gamma,\chi)$ plane as Figs.~\ref{fig:embedding_scan}(a),(b). The three regimes of the local-field case are recovered: in the \textit{transparent embedding} regime, $\chi\gtrsim10^2$, the infidelity reaches the plateau of the unembedded gadget, $1-F(J_F,\gamma)\simeq3$--$5\times10^{-3}$, and the \textit{inter-chain collapse} follows $\chi\propto\sqrt{\gamma/\omega}$, Eq.~\eqref{eq:gamma_star}. The stable region coincides with the local-field one within the resolution of the grid. At fixed dynamic range, computed from Eq.~\eqref{eq:R_range} with $J_F$ replaced by the strongest bond, $\max_A J_{F,A}=f_{\max}J_F\simeq1.8\,J_F$ for the O isotopes, the fidelity is lower than in the local-field case: $F(J_F,\gamma)\simeq0.90$--$0.92$ for $R_E=10^3$ and $F(J_F,\gamma)\simeq0.61$--$0.68$ for $R_E=10^2$, i.e.\ $0.03$--$0.04$ and $0.10$--$0.15$ below the local-field values. A global transverse field therefore introduces no additional asymptotic fitting error; the finite-resource penalty that remains consists of the additional qubits, from $N_\text{Ph}=8$ to $12$ for the O isotopes, and the per-chain programming of the bonds $J_{F,A}=f_AJ_F$.

\section{Conclusions}
\label{sec:conclusions}

In this work, we proposed a method to encode quasiparticle Hamiltonians of atomic nuclei and molecules on quantum annealers using perturbative gadgets. Through second-order Brillouin-Wigner perturbation theory, we constructed Ising Hamiltonians with local transverse fields whose low-energy spectrum reproduces the target quasiparticle Hamiltonian. We introduced two encodings: the particle-conservation encoding (PCE), which is economical in qubits but limited to a rank-one approximation of the pair hopping, and the first-quantization encoding (FQE), which requires more qubits but is systematically improvable.

We benchmarked both encodings on oxygen and calcium isotopes and on the LiH molecule, obtaining ground-state fidelities that improve systematically with the penalty strength $\gamma$ in the FQE, whereas in the PCE they saturate at the floor set by the rank-one approximation. We also simulated the adiabatic preparation of the nuclear gadget ground states, and showed how the encoding determines the structure of the energy levels and the time scale of the evolution. Neither encoding is uniformly better for the adiabatic preparation: three of the six runs reach fidelities bigger than 95\%
and the outcome follows the initial splitting of the lowest logical levels. 

Finally, we embedded the gadget Hamiltonians onto the Pegasus topology of D-Wave annealers through minor embedding and replaced the local drives with a single global transverse field, identifying the operating region of the embedding under the assumed range of hardware coefficients. The global field adds no asymptotic fitting error, at the cost of a finite overhead in qubits.

Several directions remain open. First, the gadget Hamiltonians span energy scales over several orders of magnitude, which makes them sensitive to coherent errors in the programmed couplings and fields~\cite{Pearson2019}. Incoherent effects should also be considered: at finite temperature, the logical gap must remain larger than the thermal energy after all couplings are rescaled,
which further restricts the admissible dynamic range. Second, in the regime where the gadget is most accurate,
the adiabatic protocol exceeds the coherence time of current devices~\cite{King2023}. Shortcuts to adiabaticity and optimal-control techniques, such as counterdiabatic driving, could substantially reduce the annealing time~\cite{GueryOdelin2019,Sels2017}. Third, a recently proposed annealing architecture allows initialization and measurement in arbitrary bases through auxiliary qubits~\cite{Deshpande2026}. This opens the possibility of accessing observables beyond the computational basis, while identifying which quantities relevant to nuclear physics and quantum chemistry, such as occupation numbers and pair correlations, can be extracted directly from computational-basis measurements remains a natural next step. Finally, a first hardware test could target moderate penalties just above the perturbative threshold
where the logical scale $\omega^2/\gamma$ is larger and therefore less demanding on the precision, temperature and coherence of current annealers.

Overall, perturbative gadgets turn the restricted Ising interactions of quantum annealers into a language for pairing-type many-body Hamiltonians. As annealers grow in size, coherence and control, the constructions presented here offer a concrete path to bring nuclear-structure and quantum-chemistry problems onto analog quantum hardware.

\section*{Code availability}
The Python library used to build the gadget Hamiltonians and to produce the results of this work is available at \url{https://github.com/emacosta95/Gadget4NSMForQuantumSimulators}. The numerical settings of the fits and of the time evolution are given in App.~\ref{app:numerics}.

\begin{acknowledgments}
E.C. acknowledges Cesar Fernandez-Lamas, Arnau Rios, Javier Menendez and Pablo Poggi, for the interesting discussions and insights on the project.
We used Claude Fable 5 and Claude Opus 5.5 (Anthropic) as an aid to restructure the original codes for the gadget construction and the numerical simulations into a modular, object-oriented Python library and to benchmark the numerical results, and Claude Opus 5.5 to assist in the preparation of the manuscript. All codes and results were verified independently by E.C., and both authors take full responsibility for the content of this manuscript. The authors acknowledges funding by the European Union, supported by the EuroHPC Joint Undertaking and its members under the Grant Agreement Nº 101159808, including top-up funding by the Ministry for Digital Transformation and the Civil Service of the Spanish Government. A.C.-L. also acknowledges funding from Grant RYC2022-037769-I funded by MICIU/ AEI/ 10.13039/ 501100011033 and by “ESF+". 
\end{acknowledgments}

\appendix

\section{Quasiparticle Hamiltonians Couplings}
\label{app:couplings}

\begin{figure*}[t]
    \centering
    \includegraphics[width=\textwidth]{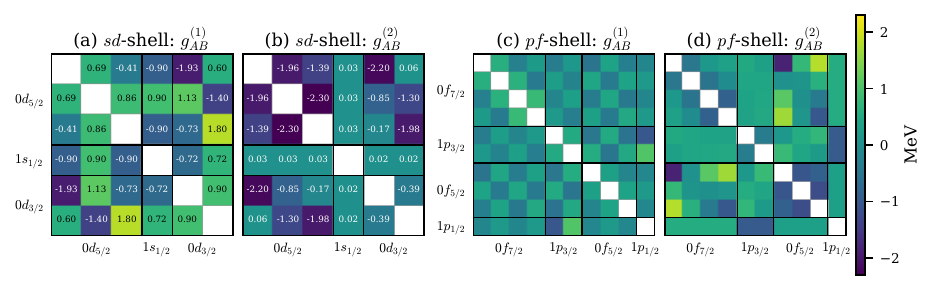}
    \caption{Couplings of $H_\text{NSM-Q}$ for neutron pairs, in MeV:
    pair-hopping amplitudes $g^{(1)}_{AB}$ and density--density couplings
    $g^{(2)}_{AB}$ for (a,b) the $sd$ shell (USDB) and (c,d) the $pf$ shell
    (GXPF1A). Rows and columns are grouped by orbital; all panels
    share the same colour scale.}
    \label{fig:couplings}
\end{figure*}

\begin{figure*}[tp]
    \centering
    \includegraphics[width=0.7\textwidth]{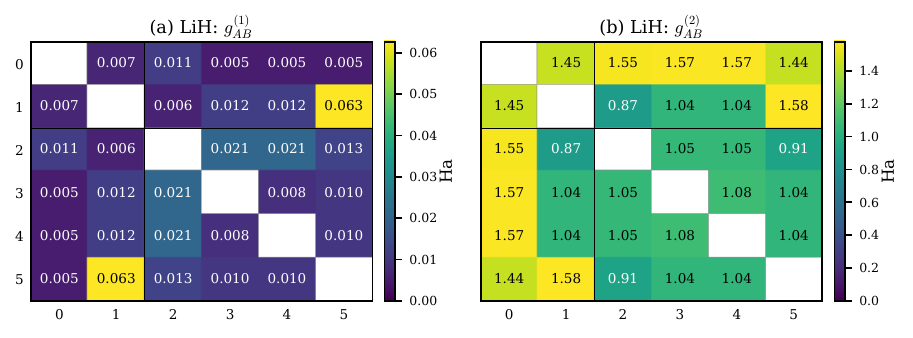}
    \caption{Couplings of $H_\text{DOCI}$ for LiH (STO-3G basis, $R=1.595$~\AA),
    in Hartree: (a) pair-hopping amplitudes $g^{(1)}_{AB}$ and
    (b) density--density couplings $g^{(2)}_{AB}$. Modes are labelled by the
    canonical RHF orbitals in order of increasing energy; the lines separate
    the occupied orbitals (0,1) from the virtual ones. Note the different
    colour scales.}
    \label{fig:couplings_LiH}
\end{figure*}

Figures~\ref{fig:couplings} and \ref{fig:couplings_LiH} show the couplings of the NSM quasiparticle Hamiltonians (Sec.~\ref{sec:nsm_qp_hamiltonian}) and of the DOCI Hamiltonian of LiH (Sec.~\ref{sec:molecular_hamiltonian}), respectively.

\section{Infidelity versus system size}
\label{app:scaling}

\begin{figure*}[tp]
    \centering
    \includegraphics[width=1\linewidth]{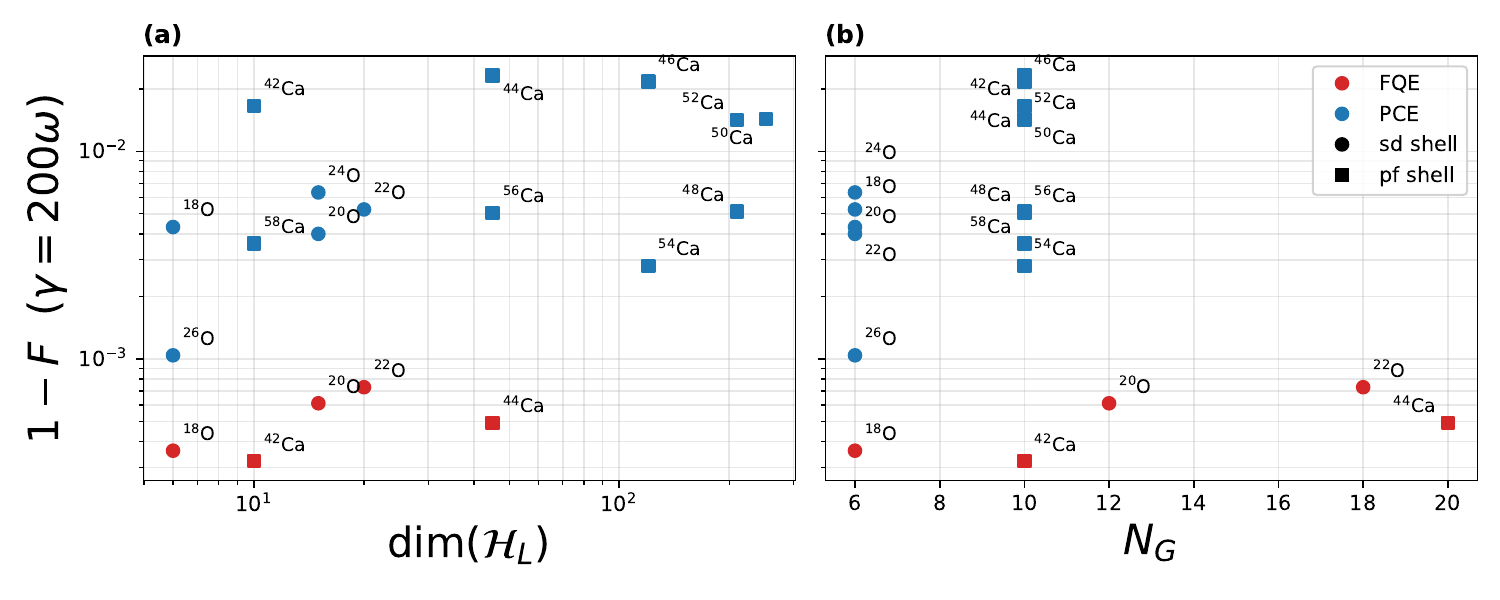}
    \caption{Infidelity at the largest penalty strength, $1-F(\gamma=200\,\omega)$, for all isotopes of Fig.~\ref{fig:4}, as a function of (a) the dimension of the quasiparticle Hilbert space, $\dim(\mathcal{H}_L)$, and (b) the number of qubits $N_G$. Red: FQE; blue: PCE. Circles: sd shell (O); squares: pf shell (Ca).}
    \label{fig:5}
\end{figure*}

Figure~\ref{fig:5} shows the infidelity at the largest penalty strength, $1-F(\gamma=200\,\omega)$, for all the NSM quasiparticle Hamiltonians of Sec.~\ref{sec:results_nsm}, as a function of (a) the dimension of the logical space, $\dim(\mathcal{H}_L)$, and (b) the number of qubits, $N_G$. For the PCE this value coincides with the plateau $1-F(\infty)$, while for the FQE it is still set by the $\gamma^{-2}$ decay and can be lowered further, down to the $k_{\max}$ fitting floor. Neither encoding shows a systematic growth of the infidelity with the system size: while $\dim(\mathcal{H}_L)$ spans more than one order of magnitude, $1-F(\gamma=200\,\omega)$ stays below $7\times10^{-4}$ for the FQE, a factor $10$--$50$ below the PCE for every nucleus, and between $10^{-3}$ and $3\times10^{-2}$ for the PCE. The spread at fixed size is as large as the one across sizes: particle-hole partners such as $^{18}$O and $^{26}$O share the same $\dim(\mathcal{H}_L)$ but differ by up to an order of magnitude in the PCE. No systematic size dependence is observed in the tested instances. In the PCE, the spread instead follows how well $g^{(1)}_{AB}$ is captured by a rank-one matrix. This also explains why the largest PCE infidelities occur in the $pf$ shell, where the rank-one ansatz has $N_q=10$ parameters for $45$ independent couplings, against $6$ for $15$ in the $sd$ shell. Panel (b) shows the price of the FQE accuracy, a qubit cost $N_G=N_\text{tot}N_q$, up to three times that of the PCE for $^{22}$O. Within the explored range, $N_G\le20$, no systematic dependence on the system size is observed at fixed $\gamma$.

\section{LiH without sign flip}
\label{app:stoq}

\begin{figure*}[tp]
    \centering
    \includegraphics[width=0.8\linewidth]{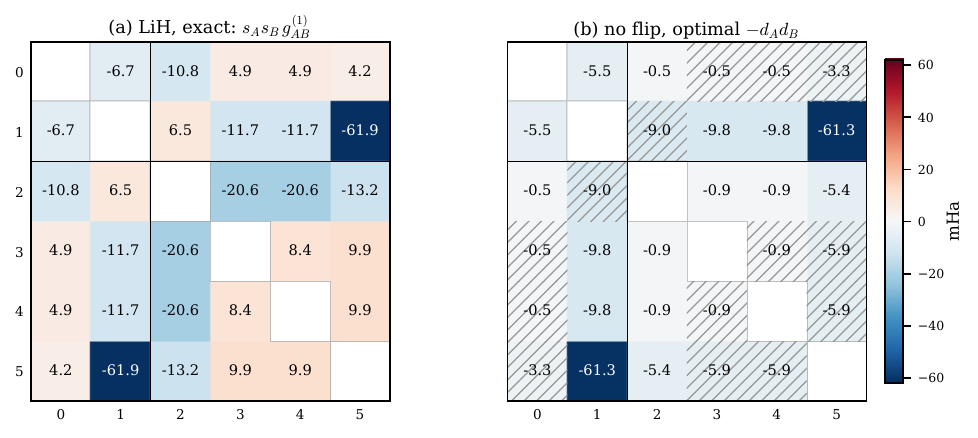}
    \caption{Pair hopping of LiH (STO-3G, canonical RHF orbitals, $R=1.595$~\AA), in mHa. (a) Exact $g^{(1)}_{AB}$ in the sign gauge $s_A=\pm1$ that minimizes the positive (frustrated) weight; black lines separate the occupied (0, 1) from the virtual (2--5) orbitals. (b) Hopping $-d_Ad_B$ generated by the PCE gadget ($k_{AB}=0$) with the optimal drive, fitted to the negative part of (a). Hatched cells: frustrated couplings, excluded from the target of the fit; the gadget still generates a nonzero hopping on them.}
    \label{fig:lih_noflip_couplings}
\end{figure*}

In Sec.~\ref{sec:results_lih} we encode $-H_\text{DOCI}$: all the hoppings become negative and the gadget matches them exactly at second order for admissible couplings, but the molecular ground state is the highest state of the logical space. Here we keep it at the bottom. Since $g^{(1)}_{AB}>0$ on every edge, every triangle of orbitals is frustrated and no choice of the orbital phases $s_A=\pm1$ makes all the couplings $s_As_Bg^{(1)}_{AB}$ negative. We take the phases that minimize the positive weight, then, to avoid the trivial solution $\textbf{d}^*=0$, we exclude from the target the couplings that remain positive, and fit the drive to the negative ones. The gadget then works in its stoquastic shape, with the target as its ground state, at the price of a inaccurate hopping. Close to equilibrium, 7 of the 15 couplings, $24\%$ of the hopping weight, are of opposite sign [Fig.~\ref{fig:lih_noflip_couplings}(a)], while the rank-one drive reproduces the dominant coupling, between the orbitals 1 and 5, within $1\%$ [Fig.~\ref{fig:lih_noflip_couplings}(b)].

The stoquastic approximation also changes the optimization of the FQE. In Sec.~\ref{sec:fqe} the drive $d^\star_A$ is fixed by the rank-one fit, and the $k_{AB}$ follow pair by pair from Eq.~\eqref{eq:k_closed}, which is exact for every pair with $\Delta_{AB}\ge1.2$. Here the excluded couplings have a zero target, which no $k_{AB}$ can reproduce: the two-stage fit would set them to $k_{\max}$, where they keep a sizable coupling set by a drive fitted at $k_{AB}=0$. We therefore minimize Eq.~\eqref{eq:loss_k} over $d_A$ and $k_{AB}$ together, with each $k_{AB}$ given at fixed $d_A$ by Eq.~\eqref{eq:k_closed}, clipped to $[0,k_{\max}]$. The lower bound replaces the condition $k_{AB}>-1$ of the main text: without an exact solution, the joint fit drifts toward the pole $k_{AB}\to-1$, where $\Delta_{AB}$ diverges and the states with two occupied qubits in the same register, at energy $(1+k_{AB})\gamma$, reach the logical space.

\begin{figure*}[tp]
    \centering
    \includegraphics[width=1\linewidth]{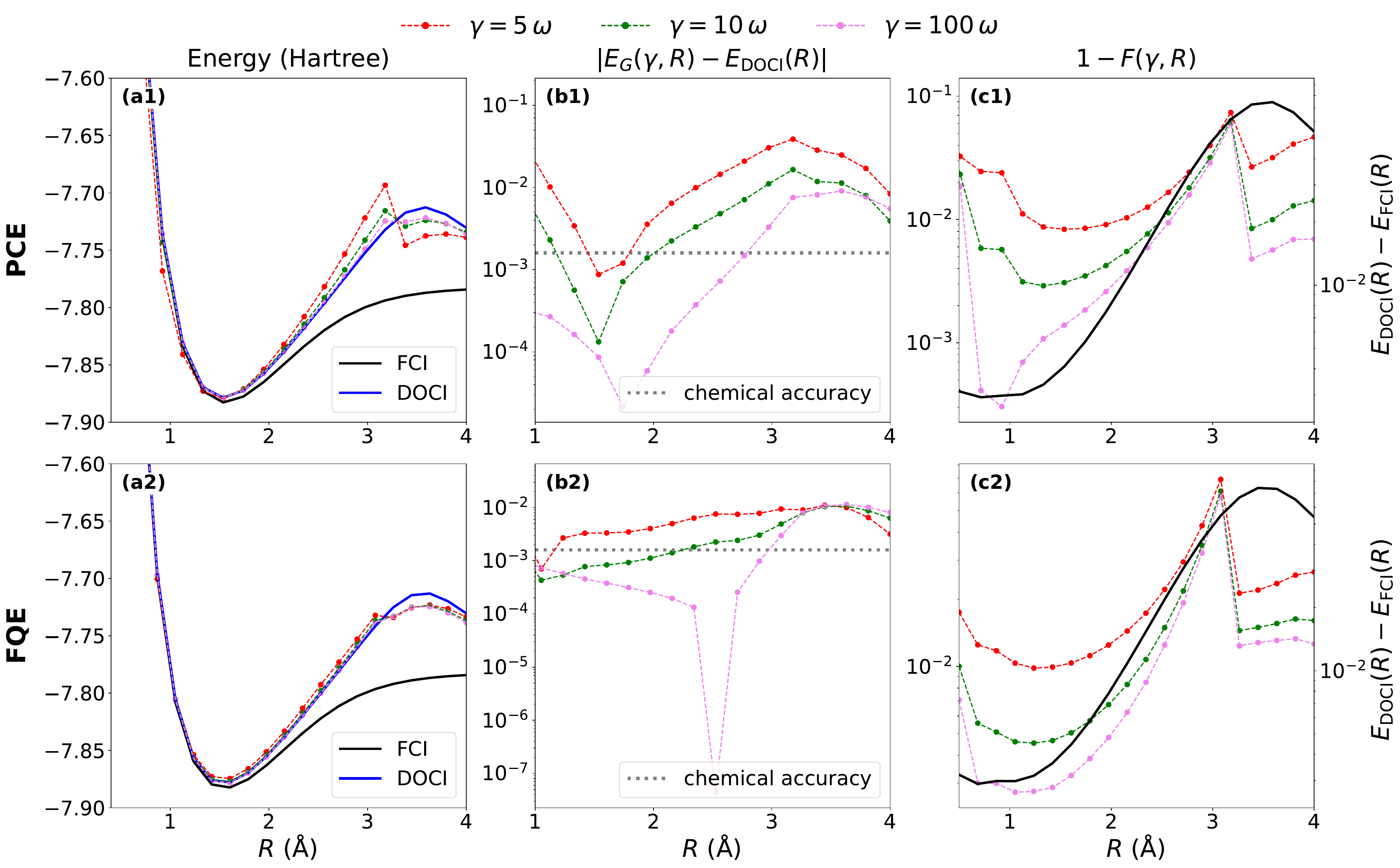}
    \caption{Gadget encoding of $H_\text{DOCI}$ for LiH without sign flip (STO-3G, canonical RHF orbitals) along the dissociation curve, in the PCE (top) and in the FQE (bottom), for $\gamma=5$, $10$ and $100\,\omega$, with $\omega=1$~Ha. (a) Ground-state energy of the gadget, compared with DOCI and FCI. (b) Energy error with respect to DOCI; dotted line: chemical accuracy, $1.6$~mHa. (c) Infidelity $1-F(\gamma,R)$ with respect to the DOCI ground state (left axis) and intrinsic error of DOCI, $E_\text{DOCI}(R)-E_\text{FCI}(R)$ (black, right axis).}
    \label{fig:lih_noflip}
\end{figure*}

Figure~\ref{fig:lih_noflip} shows that the stoquastic approximation costs a drop in accuracy. However, at $\gamma=100\,\omega$ and close to equilibrium, $1-F(\gamma,R)\simeq1.4\times10^{-3}$ in the PCE and $3\times10^{-3}$ in the FQE, with an energy error below $0.4$~mHa, an order of magnitude smaller than the intrinsic error of DOCI. The energy stays within chemical accuracy up to $R\approx2.8$~\AA\ in the PCE and $R\approx2.9$~\AA\ in the FQE.

At $R\approx3.1$--$3.2$~\AA\ both the infidelity and the energy error look discontinuous. Stretching the bond changes the relative size of the couplings, leading to a different subset of excluded couplings: the stoquastic approximation encoded by the gadget therefore changes abruptly. The infidelity grows up to the switch and drops after it, while the energy error changes sign. In this region, however, the intrinsic error of DOCI exceeds $50$~mHa, so the discontinuity has no consequence on the description of the molecule.

Compared with the sign-flip encoding of Sec.~\ref{sec:results_lih}, the stoquastic shape trades accuracy for a target that is the ground state of $H_G$, and can therefore be prepared by a forward anneal and is robust against relaxation.

\section{Effective transverse field of a ferromagnetic chain}
\label{app:chain}

We derive the prefactor of the effective transverse field in Eq.~\eqref{eq:chain_eff}. Consider a chain of $L$ physical qubits $q=0,\dots,L-1$,
\begin{equation}
    H^{\text{ch}} = -J_F\sum_{q=0}^{L-2} Z_qZ_{q+1} + V, \qquad V=\sum_{q=0}^{L-1}\Gamma_q X_q,
\end{equation}
where the longitudinal fields, $\Xi_q\ll J_F$, are neglected in the energy denominators. The ferromagnetic states $\ket{\mathbf 0}=\ket{0\cdots0}$ and $\ket{\mathbf 1}=\ket{1\cdots1}$ are degenerate. Since each term of $V$ flips one qubit, the lowest order that connects them is $L$, and each qubit is flipped exactly once. A virtual process is a trajectory labelled by the flip order $\sigma\in S_L$, and at order $L$
\begin{equation}
    \begin{split}
    \tilde d &= \bra{\mathbf 1}H_\text{eff}\ket{\mathbf 0} \\
    &= (-1)^{L-1}\Big(\prod_{q}\Gamma_q\Big)\sum_{\sigma\in S_L}\prod_{i=1}^{L-1}\frac{1}{\Delta_i(\sigma)},
    \end{split}
    \label{eq:dtilde_sum}
\end{equation}
where $\Delta_i(\sigma)$ is the excitation energy of the virtual state after the $i$-th flip, and the sign comes from the $L-1$ denominators $E_0-E_i=-\Delta_i$. The trajectories have the same order of magnitude but different weights, since $\Delta_i(\sigma)$ counts the broken bonds (kinks), each costing $2J_F$. For $L=3$, $\ket{000}\to\ket{100}\to\ket{110}\to\ket{111}$ has one kink at each step, with weight $1/(2J_F)^2$, whereas $\ket{000}\to\ket{100}\to\ket{101}\to\ket{111}$ has two kinks at the second step, with weight $1/(2J_F\cdot4J_F)$.

To make the kinks explicit, let $x_{i,q}=1$ if qubit $q$ is still in $\ket{0}$ after the $i$-th flip of trajectory $\sigma$ and $x_{i,q}=0$ otherwise. The bond $(q,q+1)$ is broken when the two qubits are in different states, so that
\begin{equation}
    \Delta_i(\sigma) = 2J_F\sum_{q=0}^{L-2}\big|x_{i,q}-x_{i,q+1}\big|.
\end{equation}
Each denominator is then written as an imaginary-time propagator, $1/\Delta=\int_0^\infty ds\,e^{-s\Delta}$, which gives
\begin{equation}
    \begin{split}
    \sum_{\sigma\in S_L}\prod_{i}\frac{1}{\Delta_i(\sigma)}
    = &\sum_{\sigma\in S_L}\int_{\mathbb{R}_+^{L-1}} ds_1\cdots ds_{L-1} \\
    &\times\exp\Big[-2J_F\sum_{i,q}s_i\big|x_{i,q}-x_{i,q+1}\big|\Big],
    \end{split}
\end{equation}
where $s_i$ is the time spent in the $i$-th virtual state. We define
\begin{equation}
    u_q = \sum_{i=1}^{L-1}s_i\,x_{i,q}\ \ge 0, \qquad q=0,\dots,L-1,
\end{equation}
the time that qubit $q$ spends in $\ket{0}$. Since the $s_i$ are positive and each qubit is flipped only once, for every pair $(q,q+1)$ one of the two sequences $x_{i,q}$, $x_{i,q+1}$ is larger than or equal to the other at every step $i$. Writing $|a-b|=\max(a,b)-\min(a,b)$, the sum over $i$ then commutes with $\max$ and $\min$, and
\begin{equation}
    \sum_{i}s_i\big|x_{i,q}-x_{i,q+1}\big| = |u_q-u_{q+1}|,
\end{equation}
the time during which the bond $(q,q+1)$ stays broken. The exponent therefore depends on $\sigma$ only through the $u_q$.

Each trajectory fixes the order of these times, since a qubit flipped later spends more time in $\ket{0}$. If $\sigma(k)$ denotes the qubit flipped at step $k$, then $u_{\sigma(k)}=s_1+\dots+s_{k-1}$ and
\begin{equation}
    u_{\sigma(L)} > u_{\sigma(L-1)} > \cdots > u_{\sigma(1)} = 0 .
    \label{eq:sector}
\end{equation}
As the exponent depends only on differences of times, we measure them with respect to qubit $0$, $\tilde u_q = u_q-u_0 = \sum_i s_i\,\tilde x_{i,q}$ with $\tilde x_{i,q}=x_{i,q}-x_{i,0}$ and $q=1,\dots,L-1$. For fixed $\sigma$, the map $(s_1,\dots,s_{L-1})\mapsto(\tilde u_1,\dots,\tilde u_{L-1})$ is linear with unit Jacobian and maps $\mathbb{R}_+^{L-1}$ onto the sector $D_\sigma\subset\mathbb{R}^{L-1}$ defined by the order of Eq.~\eqref{eq:sector}, with $\tilde u_0=0$. Different trajectories give disjoint sectors, and every order of $L$ distinct times is realised by one trajectory, so that the sum over $\sigma$ joins the sectors into the whole space, up to a set of measure zero,
\begin{equation}
    \begin{split}
    \sum_{\sigma\in S_L}\int_{\mathbb{R}_+^{L-1}} ds_1\cdots ds_{L-1}
    &= \sum_{\sigma\in S_L}\int_{D_\sigma} d\tilde u_1\cdots d\tilde u_{L-1} \\
    &= \int_{-\infty}^{+\infty} d\tilde u_1\cdots d\tilde u_{L-1}.
    \end{split}
\end{equation}
With $v_q=\tilde u_{q+1}-\tilde u_q$ ($\tilde u_0=0$), again with unit Jacobian, the integral factorizes over the bonds,
\begin{equation}
    \begin{split}
    \sum_{\sigma\in S_L}\prod_{i}\frac{1}{\Delta_i(\sigma)}
    &= \prod_{q=0}^{L-2}\int_{-\infty}^{+\infty}dv_q\,e^{-2J_F|v_q|} \\
    &= \prod_{q=0}^{L-2}2\int_0^{\infty}dv_q\,e^{-2J_Fv_q}
    = \frac{1}{J_F^{\,L-1}},
    \end{split}
\end{equation}
and Eq.~\eqref{eq:dtilde_sum} becomes
\begin{equation}
    \tilde d = (-1)^{L-1}\,\frac{\prod_q\Gamma_q}{J_F^{\,L-1}},
\end{equation}
with prefactor one. With bond-dependent couplings the same steps give $\prod_q J_{q,q+1}^{-1}$, and the result holds for chains with the shape of a tree, whose $L-1$ bond differences are again independent variables.

\subsection{Virtual processes and the breakdown at \texorpdfstring{$\gamma^*$}{gamma*}}
\label{app:asym}

Consider a chain $A$ with $L_A=2$, as in the embeddings of Fig.~\ref{fig:embedding_scan}, and a basis state $\ket{S}$ of the logical space, i.e., a pair Slater determinant with occupation numbers $S=(n_1,\dots,n_{N_G})$, where $n_B$ is the eigenvalue of $N_B$. In $\ket{S}$ all the chains are frozen. Flipping one qubit $q=1,2$ of chain $A$ breaks the bond and costs $2J_F+\delta_q(A,S)$, where
\begin{equation}
    \delta_q(A,S)=\big(1-2n_A\big)\Big[\frac{\Xi_A}{2}+\sum_{B \in S/A}n_B\,\frac{\Lambda_{AB}}{|\mathcal{E}_{AB}|}\,e^{AB}_q\Big]
    \label{eq:delta_q}
\end{equation}
is the change of the longitudinal energy of Eq.~\eqref{eq:HG_PC_emb} produced by the flip. $e^{AB}_q$ is the number of couplers of $\mathcal{E}_{AB}$ attached to the flipped qubit $q$. The field $\Xi_A$, the couplings $\Lambda_{AB}$, and the numbers $e^{AB}_q$ are fixed by the Hamiltonian and by the embedding. Flipping both qubits takes the gadget qubit out of the logical space and costs the penalty $\gamma$, so $\delta_1(A,S)+\delta_2(A,S)=\gamma+O(\omega^2/\gamma)$ for every $S$, where the correction comes from the target diagonal terms in $\Xi_A$ and $\Lambda_{AB}$. Individually, however, the $\delta_q(A,S)$ are of order $\gamma$ and depend on $S$, because the inter-chain terms $N_{A,r}N_{B,s}$ of Eq.~\eqref{eq:HG_PC_emb} act only on the qubits attached to the occupied chains $B$.

The presence of $\delta_{1,2}(A,S)$ in the virtual processes modifies both the transverse field and the energy of the frozen configuration. Summing the two virtual paths, through $\ket{10}$ and $\ket{01}$, that connect $\ket{00}$ (energy $0$) and $\ket{11}$ (energy $\gamma$), at second order in $\Gamma_A$, with $\Gamma_A^2=J_F|d_A|$ [Eq.~\eqref{eq:chain_fields}], gives
\begin{equation}
    \begin{split}
    \tilde d_A'(S)&=\frac{\Gamma_A^2}{2}\sum_{q=1,2}\left[\frac{1}{-2J_F-\delta_q}+\frac{1}{\gamma-2J_F-\delta_q}\right]\\
    &=-\frac{\Gamma_A^2}{J_F}\left[1+\frac{\delta_1^2+\delta_2^2}{8J_F^2}+\mathcal{O}\!\left(\chi^{-4}\right)\right],
    \end{split}
    \label{eq:dtilde_asym}
\end{equation}
where $\delta_q\equiv\delta_q(A,S)$; since $\gamma-\delta_1=\delta_2 + O(\omega^2/\gamma)$, the two denominators are $2J_F\pm\delta_q$, so all the odd powers of $\delta_q$ cancel and the first neglected term is of order $\chi^{-4}$. The same virtual states also generate an additional effective longitudinal field, i.e., an energy shift of the frozen configuration,
\begin{equation}
    \begin{split}
    \tilde h_A'(S)&=-\Gamma_A^2\sum_{q=1,2}\frac{1}{2J_F+\delta_q}\\
    &=-|d_A|\left[1-\frac{1}{4\chi}+\frac{\delta_1^2+\delta_2^2}{8J_F^2}+\mathcal{O}\!\left(\chi^{-3}\right)\right].
    \end{split}
    \label{eq:sigma_asym}
\end{equation}
Here, unlike in Eq.~\eqref{eq:dtilde_asym}, each virtual path returns to the same state and has a single energy denominator, so all powers of $\delta_q$ survive and the first neglected term is of order $\chi^{-3}$. The first order term $1/\chi$ does not depend on $A$ and $S$, and therefore it acts as a global shift in the energy spectrum. The second-order term, instead, depends on $S$ through $\delta_1(A,S)^2+\delta_2(A,S)^2$, so that the effective longitudinal field is no longer uniform over the logical states. Since $\delta_q\sim\gamma$, both corrections are of order $(\gamma/J_F)^2=1/\chi^2$ and depend on $S$ through $\delta_q(A,S)$: the deviation of the transverse field is $\Delta\tilde d_A\sim d_A/\chi^2$, and the deviation of the effective longitudinal field is $\Delta\tilde h_A\sim|d_A|/\chi^2$. The embedded chains therefore add to the effective Hamiltonian $H_\text{eff}$ an extra perturbation $\Delta H$ of order
\begin{equation}
    \|\Delta H\|\sim\frac{|d_A|}{\chi^2}.
    \label{eq:DeltaH}
\end{equation}

To see how $\Delta H$ affects the fidelity, we restrict $H_\text{eff}+\Delta H$ to the ground state $\ket{\Psi_0}$ of the effective Hamiltonian $H_\text{eff}$, with energy $E_0$, and to the excited state $\ket{\Psi_\perp}$, with energy $E_\perp$, that is closest in energy among those coupled to $\ket{\Psi_0}$ by $\Delta H$. In this basis
\begin{equation}
    H_\text{eff}+\Delta H\simeq
    \begin{pmatrix}
    E_0+\Delta H_{00} & \Delta\\
    \Delta^* & E_\perp+\Delta H_{\perp\perp}
    \end{pmatrix},
    \label{eq:two_level}
\end{equation}
with $\Delta H_{00}=\braket{\Psi_0|\Delta H|\Psi_0}$, $\Delta H_{\perp\perp}=\braket{\Psi_\perp|\Delta H|\Psi_\perp}$, and $\Delta=\braket{\Psi_0|\Delta H|\Psi_\perp}$. Since $\Delta H\sim|d_A|/\chi^2$, all three matrix elements are of this order. The two levels are separated by the effective gap
\begin{equation}
    \begin{split}
    D&=\Delta_L-\left(\Delta H_{00}-\Delta H_{\perp\perp}\right),\\
    \Delta_L&=E_\perp-E_0=\frac{\omega \;\Delta_G}{\gamma},
    \end{split}
\end{equation}
where $\Delta_G$ is the gap of $(\gamma/\omega) H_\text{eff}$, i.e., of the effective Hamiltonian generated by the gadget without the embedding, which coincides with the gap $\Delta_\text{qp}$ of the quasiparticle Hamiltonian up to the error of the gadget encoding. The eigenvalues of Eq.~\eqref{eq:two_level} are, exactly,
\begin{equation}
    \lambda_\pm=\frac{E_0+E_\perp+\Delta H_{00}+\Delta H_{\perp\perp}}{2}\pm\frac{\sqrt{D^2+4|\Delta|^2}}{2}.
\end{equation}
The ground state of $H_\text{eff}+\Delta H$ is the eigenvector of $\lambda_-$,
\begin{equation}
    \ket{\Psi'_0}=\frac{\Delta\ket{\Psi_0}+(\lambda_--E_0-\Delta H_{00})\ket{\Psi_\perp}}{\sqrt{|\Delta|^2+(\lambda_--E_0-\Delta H_{00})^2}},
\end{equation}
and its weight on $\ket{\Psi_\perp}$ defines the deviation $\delta F\equiv|\braket{\Psi_\perp|\Psi'_0}|^2=1-|\braket{\Psi_0|\Psi'_0}|^2$, i.e., the infidelity added by the embedding with respect to the ground state of the unembedded gadget:
\begin{equation}
    \delta F=\frac12\left[1-\frac{D}{\sqrt{D^2+4|\Delta|^2}}\right],
    \label{eq:F_two_level}
\end{equation}
which is exact within the two-level model. Far from the breakdown, $D\gg|\Delta|$, and to leading order in $|\Delta|/D$
\begin{equation}
    \delta F\simeq\frac{|\Delta|^2}{D^2}\sim\left(\frac{\gamma\,|d_A|}{\chi^2\,\Delta_G}\right)^2.
    \label{eq:dF_scaling}
\end{equation}
This regime ends when $|\Delta|\sim D$, that is, for
\begin{equation}
    \gamma^*\sim\frac{\Delta_G}{|d_A|}\,\chi^2
    \quad\Longleftrightarrow\quad
    \gamma^*\sim\left(\frac{\Delta_G}{|d_A|}\right)^{1/3}J_F^{2/3},
    \label{eq:gamma_star_app}
\end{equation}
which reproduces the scaling of Eq.~\eqref{eq:gamma_star}. This derivation assumes $L_A=2$, as for all the chains of our embeddings. For longer chains the transverse fields $\Gamma_{A,r}$ scale differently with $J_F$, and, depending on the position of the couplers along the chain, the term linear in $\delta_q$ may no longer cancel. As far as we can tell, the breakdown still follows a power law $\gamma^*\propto\chi^p$, but with an exponent $p$ that depends on the length of the chains.

\section{Numerical details of the simulations}
\label{app:numerics}

All the fits, diagonalizations, and time evolutions are performed with SciPy, whereas the many-body operators are built with QuTiP~\cite{Johansson2012,Johansson2013,Lambert2026}.

\textit{Rank-one fit.} The loss of Eq.~\eqref{eq:loss_rank1} is minimized with L-BFGS-B, with analytic gradient, and at most $10^4$ iterations. The runs starts with random initializations, and the run with the lowest loss is kept. 

\textit{FQE fits.} In the two-stage fit of Sec.~\ref{sec:fqe} the $k_{AB}$ follow from Eq.~\eqref{eq:k_closed} at fixed $d_A$, and the pairs with $k_{AB}\le-1$ or $k_{AB}>k_{\max}$ are set to $k_{\max}$. In the joint fit, the residuals of Eq.~\eqref{eq:loss_k} are minimized over $d_A$, with each $k_{AB}$ given at the current $d_A$ by Eq.~\eqref{eq:k_closed} and clipped to $[k_{\min},k_{\max}]$. The runs start from the rank-one drive and from 150 random drives, and the run with the smallest maximum residual is kept. 

\textit{Spectra and dynamics.} Spectra and static fidelities are obtained by exact diagonalization defining the Hamiltonian matrices in sparse rappresentation using Qutip. The annealing dynamics of Sec.~\ref{sec:qa} is computed with a trotterized unitary dynamics: $H(t)$ is held fixed over steps of length $\delta t=0.1\,\omega^{-1}$, and each step, $e^{-iH(t_k)\delta t}$, is applied to the state with the Krylov-type algorithm of SciPy's \texttt{expm\_multiply}~\cite{AlMohy2011}.

\bibliographystyle{apsrev4-2}
\bibliography{references}

\end{document}